\documentclass[journal]{IEEEtran}
\usepackage{amsmath}
\usepackage{amsfonts}
\usepackage{amsthm}
\usepackage{cite}
\usepackage{soul,color}
\usepackage{graphicx}
\usepackage{float}
\usepackage[font=footnotesize]{caption}
\usepackage[font=footnotesize]{subcaption}
\usepackage{epstopdf}
\usepackage{amssymb}
\usepackage{stmaryrd}
\usepackage{multirow}

\usepackage{enumitem}
\usepackage{algorithm}
\usepackage{algorithmic}
\begin{document} 
\title{Invoking Deep Learning for Joint Estimation of Indoor LiFi User
Position and Orientation}
\author{Mohamed Amine Arfaoui$^{*}$,
        Mohammad Dehghani Soltani,
        Iman Tavakkolnia,
        Ali Ghrayeb, 
        Chadi Assi, \\
        Majid Safari, 
        and Harald Haas \vspace{-0.7cm}
\thanks{M. A. Arfaoui and C. Assi are with Concordia Institute for Information Systems Engineering (CIISE), Concordia University, Montreal, Canada, e-mail:\{m\_arfaou@encs, assi@ciise\}.concordia.ca.}
\thanks{M. D. Soltani, I. Tavakkolnia, M. Safari, and H. Haas are with the LiFi Research and Development Centre, Institute for Digital Communications, School of Engineering, University of Edinburgh, UK. e-mail: \{m.dehghani, i.tavakkolnia, majid.safari, h.haas\}@ed.ac.uk.}
\thanks{A. Ghrayeb is with the Electrical and Computer Engineering (ECE) department, Texas A$\&$M University at Qatar, Doha, Qatar, e-mail: ali.ghrayeb@qatar.tamu.edu.}
\thanks{$^*$\textit{Corresponding author: M.A. Arfaoui, m$\_$arfaou@encs.concordia.ca}}
}
\maketitle 
\thispagestyle{plain}
\begin{abstract} Light-fidelity (LiFi) is a fully-networked bidirectional optical wireless communication (OWC) technology that is considered as a promising solution for high-speed indoor connectivity. In this paper, the joint estimation of user 3D position and user equipment (UE) orientation in indoor LiFi systems with unknown emission power is investigated. Existing solutions for this problem assume either ideal LiFi system settings or perfect knowledge of the UE states, rendering them unsuitable for realistic LiFi systems. In addition, these solutions consider the non-line-of-sight (NLOS) links of the LiFi channel gain as a source of deterioration for the estimation performance instead of harnessing these components in improving the position and the orientation estimation performance. This is mainly due to the lack of appropriate estimation techniques that can extract the position and orientation information hidden in these components. In this paper, and against the above limitations, the UE is assumed to be connected with at least one access point (AP), i.e., at least one active LiFi link. Fingerprinting is employed as an estimation technique and the received signal-to-noise ratio (SNR) is used as an estimation metric, where both the line-of-sight (LOS) and NLOS components of the LiFi channel are considered. Motivated by the success of deep learning techniques in solving several complex estimation and prediction problems, we employ two deep artificial neural network (ANN) models, one based on the multilayer perceptron (MLP) and the second on the convolutional neural network (CNN), that can map efficiently the instantaneous received SNR with the user 3D position and the UE orientation. Through numerous examples, we investigate the performance of the proposed schemes in terms of the average estimation error, precision, computational time, and the bit error rate. We also compare this performance to that of the k-nearest neighbours (KNN) scheme, which is widely used in solving wireless localization problems. It is demonstrated that the proposed schemes achieve significant gains and are superior to the KNN scheme. 
\end{abstract} 
\IEEEpeerreviewmaketitle 
\section{Introduction}
\indent With the dramatic increase in the data traffic, fifth generation (5G) and beyond networks must urgently provide high data rates, seamless connectivity, ubiquitous coverage and ultra-low latency communications \cite{Intro2,Intro3}. In addition, with the emergence of the Internet of-Things (IoTs) networks, the number of connected devices to the internet is increasing dramatically \cite{Intro6}. This fact implies not only a significant increase in data traffic, but also the emergence of IoT services with crucial requirements, such as higher data rates, higher connection density, ultra reliable and low latency communication (URLLC) \cite{intro5}. Nowadays, the availability of location and data of mobile terminals at the communications stations (access points (APs) and base stations (BSs)), i.e., their knowledge by the telecommunications operators, has become a key factor in enabling next generation communication systems. Such information enables better estimation of the quality of the wireless links, which can improve the resource management and provide new location-based services \cite{yassin2016recent}. \\ 
\indent Global Navigation Satellite Systems (GNSS), such as Global Positioning Systems (GPS), and the standalone cellular systems are the present mainstream in positioning systems, and they are widely used in aircraft, vehicles, and portable devices in order to provide real-time positioning and navigation \cite{kaplan2005understanding}. However, in indoor environments, these positioning systems are severely degraded or may fail altogether since the signals transmitted by the satellite or the cellular networks are usually degraded and interrupted by clouds, ceilings, walls, and other obstructions \cite{lan2015novel,zhuang2015tightly}. On the other hand, indoor applications require much more accurate positioning than outdoor applications. Consequently, indoor positioning systems using indoor wireless signals (e.g., wireless-fidelity (WiFi) \cite{zhuang2015evaluation}, Bluetooth \cite{zhuang2016smartphone}, radio frequency identification (RFID) \cite{yang2014efficient}, and ZigBee \cite{fang2012enhanced}) have been proposed to fill the gap of GPS and cellular signals to improve the performance of indoor positioning.  \\ 
\indent The knowledge of user equipment (UE) position and orientation is a crucial factor for indoor location-based applications such as robotic navigation \cite{yao2018optimizing} and autonomous parcel sorting \cite{eppner2016lessons}. Although WiFi and Bluetooth are the most utilized positioning systems, which have already been widely deployed in current smart devices, they cannot satisfy the requirements (UE position and orientation) of the above applications since their localization performance suffers from the limited number of available APs in their local area \cite{zhou2016particle}. Due to this issue, novel and accurate position and orientation estimation solutions are highly demanded.
\subsection{LiFi-Based Indoor Positioning Systems}
\indent Light-fidelity (LiFi) is a novel bidirectional, high speed and fully networked wireless communication technology, that uses visible light as the propagation medium in the downlink for the purposes of illumination and communication \cite{haas2015lifi}. It uses infrared (IR) in the uplink so that the illumination constraint of a room remains unaffected, and also to avoid interference with the visible light in the downlink \cite{haas2015lifi}. LiFi offers a number of important benefits that have made it favorable for future technologies. These include the very large, unregulated bandwidth available in the visible light spectrum, high energy efficiency \cite{tavakkolnia2018energy}, the straightforward deployment that uses off-the-shelf light emitting diodes (LEDs) and photodiode (PD) devices at the transmitter and receiver ends, respectively, and enhanced security as light does not penetrate through walls and opaque objects \cite{arfaoui2020physical}. \\
\indent It is predictable that implementing a novel localization technology based on LiFi systems has a great potential, which has encouraged both academia and industry to step into the field \cite{qualcommlumicast}. However, it is worth mentioning that the device orientation is a crucial factor in LiFi networks as well as mmWave systems \cite{Zhihong_VTCfall_2018}, which are two possible choices for future indoor communications that can fulfil high-data-rate requirements of users. Hence, reporting the device orientation along with the position can remarkably help to improve user quality of service, resource allocation, and interference management for these networks \cite{MDS2019Thesis}. \\
\indent Over the past few years, many algorithms for LiFi-based indoor positioning have been proposed and verified by experiments. LiFi-based indoor positioning systems have shown to be more accurate (0.1-0.35 m positioning error) when compared to WiFi (1-7 m), Bluetooth (2-5 m), and other technologies \cite{hassan2015indoor}. RF-based positioning metrics and algorithms have been developed for indoor positioning systems and these are also applicable to LiFi-based positioning systems. In fact, we distinguish between three main positioning metrics, namely, the received signal strength (RSS), the time of arrival (TOA) and the angle of arrival (AOA) \cite{yassin2016recent}. The RSS measures the power of the received signals, which follows the channel model in general and hence the position estimation can be obtained. The TOA measures the travel time of the signal from the transmitter to the receiver, which is a function of the distance as well. Finally, the AOA measures the angle from which the signal arrives at the receiver and such information can be also exploited in estimating the location of the transmitter \cite{yassin2016recent}. 
\subsection{Existing Solutions}
Several LiFi-based indoor positioning solutions have been proposed in the literature \cite{yang2014three,sharifi2016indoor,zhang2014asynchronous,zhou2012indoor,yin2015indoor,qiu2015visible}. An AOA-based technique is proposed in \cite{yang2014three}, which uses a receiver array with known orientation angle differences between receivers.  In \cite{sharifi2016indoor,zhang2014asynchronous,zhou2012indoor} both the LED transmitters and the UE receiver are assumed to have perpendicular orientations to the room ceiling, and the height of the UE is assumed to be known. However, the assumption of having a perfect alignment between the UE orientation and the LED transmitter orientation is not valid in real-life scenarios. In fact, the majority of studies on LiFi systems assume that the device is always perfectly aligned to the APs. This assumption may have been driven by the lack of having a proper model for orientation, and/or to make the analysis tractable. Such an assumption is only accurate for a limited number of devices (e.g., laptops with a LiFi dongle), while the majority of users use devices such as smartphones, and in real-life scenarios, users tend to hold their devices in a way that feel most comfortable. Due to this, an inertial measurement unit (IMU) was required in \cite{yin2015indoor} to measure the UE tilt angle for position estimation. However, the IMU may not be available in some real-life scenarios and the accuracy of the IMU in estimating the UE tilt angle is not also guaranteed. \\
\indent Based on the above discussion, there are a number of limitations in the aforementioned approaches. In fact, only the estimate of the UE location was considered and hence the estimate of the UE orientation remains unresolved.  This is mainly due to the fact that the position and orientation estimation metrics of LiFi systems, such as the RSS, are non-linear functions with respect to (w.r.t.) the UE position and orientation \cite{soltani2018modeling}, which leads to a non-convex optimization problem with a lot of local optima. In addition, the imperfect estimation of these uncertain parameters will result in a serious estimation performance loss. Despite this, the orientation estimation in LiFi systems should not be ignored. In fact, unlike conventional radio frequency wireless systems, the LiFi channel is not isotropic, meaning that the device orientation affects the channel gain significantly, which makes the orientation estimation a crucial factor. In addition, such orientation can affect the users' bit error rate (BER) and throughput remarkably and it should be estimated carefully \cite{soltani2018modeling,purwita2018WCNC,Zhihong_VTCfall_2018,MDS2019Thesis}.\\ 
\indent In \cite{zhou2019joint}, a simultaneous position and orientation (SPAO) algorithm for indoor LiFi users with unknown LED emission power is proposed. This approach is based on the RSS where an iterative algorithm for jointly estimating the UE position and orientation is developed. Although the proposed approach considers estimating the UE orientation, it does require that the UE should be connected to at least six APs, which is equal to the number of unknown parameters (three parameters for the UE position and three others for the UE orientation). However, such an assumption on the system setting is not valid in realistic LiFi systems due to the random orientation of LiFi UEs. Against the above background, an accurate LiFi-based indoor position and orientation estimation solution without any requirement regarding the LiFi system settings or perfect knowledge of the UE states is highly desirable. \\ 
\indent The existing position and orientation estimation solutions discussed above considered only the line of sight (LOS) component in estimating the UE position and orientation and treated the non-line-of-sight (NLOS) components as a source of noise that deteriorates the estimation performance. This is mainly due to the fact that the expression of the NLOS channel gain w.r.t the UE location and orientation is complex, and hence, it could not be handled straightforwardly in an optimization fashion. However, it was shown recently in \cite{zhou2020performance} that LiFi systems can gain additional UE position and orientation information from the NLOS links via leveraging the NLOS propagation knowledge. Specifically, the closed-form Cramer-Raw lower bounds (CRLBs) on the estimation errors of the UE location and orientation, are derived. In addition, the information contribution of NLOS links was quantified to gain insights into the effect of NLOS propagation on the LiFi-based indoor position and orientation estimation performance. It was shown that the NLOS channel, in addition to the LOS channel, can be exploited to improve the LiFi-based indoor position and orientation estimation performance. However, due to the fact that the channel gain expression of the NLOS components is very complex w.r.t. the UE position and orientation, including the NLOS components in the estimation process is not straightforward from an optimization point of view. Therefore, this gives rise to the following question: \textit{"How can the NLOS components be exploited in estimating the UE location and orientation in LiFi-based indoor environments?"} The answer is in fact using deep learning (DL) techniques.
\subsection{The Need for Deep Learning}
\indent As a prevailing approach to artificial intelligence, machine learning (ML) has drawn much attention in recent years due to its great successes in computer vision and natural language processing \cite{wang2017deep}. ML is capable of solving complex problems that are lacking explicit models or straightforward programming. Motivated by its successful applications to many practical tasks, both industry and the research communities have advocated the applications of ML in wireless communication, with emphasis on resource management, networking, mobility management and localization \cite{zhu2020toward,sun2019application}. Recently, some works have investigated the use of ML techniques in indoor positioning using LiFi technology, such as K-Nearest Neighbor (KNN) \cite{tran2020high}, support vector machine (SVM) and extreme learning machine (ELM) \cite{chen2019indoor,wang2019machine}. \\
\indent DL is a particular ML technique that implements the learning process elaborating the data through ANNs. The use of ANNs is a key factor that makes DL outperform other machine learning schemes, especially when a large amount of data is available \cite{zappone2019wireless}. This has made DL the leading ML technique in many scientific fields such as image classification, text recognition, speech recognition, audio and language processing and robotics \cite{zappone2019wireless}. The potential application of DL to physical layer communications has also been increasingly recognized because of the new features for future communications, such as complex scenarios with unknown channel models, high speed and accurate processing requirements, which present big challenges to 5G and beyond wireless networks \cite{wang2017deep}. Motivated by this, DL has been applied to wireless communications, such as physical layer communications \cite{wang2017deep,qin2019deep}, resource allocation \cite{ye2019deep,tang2018novel}, and intelligent traffic control \cite{tang2017removing}. Motivated by the above discussion, DL techniques are auspicious candidates for LiFi-based indoor position and orientation estimation. Therefore, the use of ANNs is a promising solution for this problem, which is the focus of this paper.
\subsection{Contributions and Outcomes}
\indent In this paper, a UE with a random orientation is located randomly within an indoor environment. The UE communicates on the uplink channel (using IR signals) with multiple APs mounted on the ceiling of the indoor environment. The objective is the simultaneous 3D-position and orientation estimation of a LiFi-enabled UE. Unlike existing positioning methods, the proposed scheme does not require specific settings of the LiFi system such as the number of active links with the APs, or having prior knowledge about the UE position and orientation, or the emitting power of the positioning signals transmitted by the UE. In fact, we only assume that the UE is expected to be connected to at least one AP in order to be within the coverage of the LiFi system. \\ 
\indent The adopted position and orientation estimation approach is the RSS-based fingerprinting and it consists of two different phases, an offline phase and an online phase, In the offline phase, the received SNR at the APs for possible 3D positions and orientation angles are collected, processed and recorded into a dataset. Then, based on the obtained dataset, two ANNs are proposed for mapping the instantaneous received SNR values with the 3D position and orientation, where the first is based on the multiple layer perceptron (MLP) and the second is based on the convolutional neural network (CNN). In the online phase, the ANN models are deployed. The APs receive signals from the UE with unknown positions, orientations, and emission power. Then, it applies the trained ANN models to the SNR values and estimates the 3D position and the orientation of the UE. \\
\indent The performance of the proposed models, in terms of estimation error, precision, computational time and the BER, is compared with the KNN approach, which is the best ML-based fingerprinting approach in LiFi positioning systems reported in the literature \cite{poulose2020performance,tran2020high}. We demonstrate the superiority of the proposed models through several examples. In addition, the gain obtained by including the NLOS components in the estimation methods is evaluated and the effect of the size of the dataset on the estimation performance is also investigated. The simulation results show that the CNN model is the most accurate model. In fact, for an indoor environment of size $5\times 5\times 3$ m$^3$ and for a dataset of size $10^5$, the proposed CNN model is able to achieve an average positioning error of $16.49$ cm when considering the LOS and NLOS components of the channel gain and of $21.83$ cm when considering only the LOS components. However, for the same indoor environment and for a dataset of size $10^6$, the proposed CNN model is able to achieve an average positioning error of $10.53$ cm when considering the LOS and NLOS components of the channel gain and of $14.55$ cm when considering only the LOS components. In addition, the average estimation time required for the proposed CNN model in the online phase is about $18$ millisecond (ms), which demonstrates the potential of the proposed approach in providing extremely fast and accurate estimation. Finally, based on these observations, some future research directions are proposed.
\subsection{Outline and Notation}
\indent The rest of the paper is organized as follows. The system model is presented in Section II. Section III presents the proposed RSS-based fingerprinting approach. Sections IV presents the simulation results. Finally, the paper is concluded in Section V and future research directions are highlighted. \\ 
\indent The notations adopted throughout the paper are summarized in Table \ref{T1}. In addition, for every random variable $X$, $f_X$ denotes the probability density function (PDF) of $X$. $\mathbb{E} \left[ \cdot \right]$ denotes the expected value. The function $\mathcal{U}_{[a,b]}\left( \cdot \right)$ denotes the unit step function within $[a,b]$, i.e., for all $x \in \mathbb{R}$, $\mathcal{U}_{[a,b]}\left( x \right) = 1$ if $x \in [a,b]$, and 0 otherwise. Finally, for $N \in \mathbb{N}$, $\textbf{0}_N$ and $\textbf{1}_N$ denote the all zeros and all ones $N \times 1$ vector, respectively.
\begin{table}[t]
\centering
\caption{Table of Notations}
\renewcommand{\arraystretch}{1} 
\setlength{\tabcolsep}{0.2cm} 
\begin{tabular}{| l | l |}
  \hline 
  \multicolumn{2}{|c|}{\textbf{Environment Geometry}} \\
  \hline 
  \hline
  $L \times W \times H$ & Dimension of the indoor environment \\ 
  \hline
  $N_{\rm r}$ & Number of the APs \\ 
  \hline
  $N_{\rm t}$ & Number of IR-LEDs of the UE \\ 
  \hline
  $K$ & Number of surface elements of the indoor \\
   & environment \\ 
  \hline
  $\zeta_k$ & Reflectivity coefficients of the indoor environment \\ 
  \hline 
  \hline 
  \multicolumn{2}{|c|}{\textbf{LiFi Channel Parameters}} \\
  \hline 
  \hline 
  $\textbf{H}$ & Channel matrix\\ 
  \hline 
  $\boldsymbol{\rho}$ & Received SNR \\
  \hline 
  $(x,y,z)$ & UE 3D position \\ 
  \hline 
  $(\alpha, \beta, \gamma)$ & UE orientation angles \\ 
  \hline 
  $P_{\rm elec}$ & Transmitted electrical power \\
  \hline 
\end{tabular} 
\label{T1}
\end{table}
\section{System Model} 
\subsection{System Setup}
\begin{figure}[t]
\centering     
\includegraphics[width=1\linewidth]{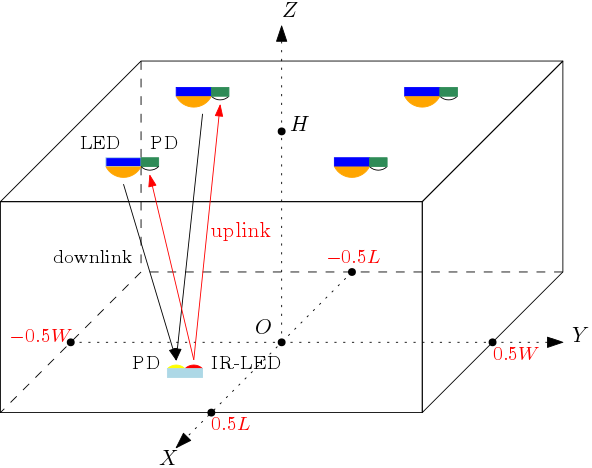}
\caption{A typical indoor LiFi system.}
\label{fig:IndEnv}
\end{figure}

We consider the indoor LiFi system shown in Fig.~\ref{fig:IndEnv}, which consists of a room with size $L\times W \times H$, where $L$, $W$ and $H$ denote the length, the width and the height of the room, respectively. The LiFi system is equipped with $N_{\rm r}$ APs installed at the ceiling of the room. Each AP is down facing and is equipped with one LED and one PD adjacent to each other, where the LED is used for illumination and data transmission simultaneously and the PD is used for data reception. In addition, a LiFi-enabled UE is randomly located within the room and it is equipped with $N_{\rm t}$  infrared LEDs (IR-LEDs) and $N_{\rm t}$ PDs that are used for data transmission and reception, respectively. The IR-LEDs and PDs of the UE are grouped into $N_{\rm t}$ pairs, where each pair consists of one IR-LED and one PD that are adjacent to each other. As shown in Fig.~\ref{fig:IndEnv}, the communication between the APs and the UE is bidirectional. Specifically, in the downlink, the APs employ the visible light spectrum for transmitting the information and the LiFi UE receives this information through its PDs, whereas in the uplink, the IR-LEDs of the UE transmit the information using the IR spectrum and the APs detect the transmitted signals through their PDs. In this mechanism, there is no interference between the downlink and uplink transmissions and the two phases can occur simultaneously. 
\subsection{Transmission Model}
\indent In the system model, we focus on the uplink transmission, where the intensity modulation direct detection (IM/DD) is considered. The transmit elements are the $N_{\rm t}$ IR-LEDs of the UE that are driven by a fixed bias current $I_{\rm DC} \in \mathbb{R}_{+}$, which sets the average radiated optical power from the IR-LEDs. The data signals are grouped into an $N_{\rm t} \times 1$ zero-mean vector of current signals $\textbf{s}$, which is then superimposed on $I_{\rm DC}$, via, e.g., a bias-T circuit, to imperceptibly modulate the instantaneous optical power $\textbf{P}_{\rm t}$ emitted from the IR-LEDs. Then, using an appropriate pre-distorter \cite{elgala2009predistortion}, the electro-optical conversion can be modeled as $\textbf{P}_{\rm t} = \eta \left(\textbf{s} + I_{\rm DC}\textbf{1}_{N_{\rm t}} \right)$, where $\eta$ (W/A) is the current-to-power conversion efficiency of the IR-LEDs. Since $\mathbb{E}\left(\textbf{s}\right)=0$, the data signals do not contribute to the average optical power. The optical powers collected by the PDs of the APs are given by $\textbf{P}_{\rm r} = \textbf{H} \textbf{P}_{\rm t}$, where $\textbf{H}$ is the $N_{\rm r} \times N_{\rm t}$ channel matrix between the $N_{\rm t}$ IR-LEDs and the $N_{\rm r}$ PDs of the APs. The PDs of the APs, with responsivity $R_{\rm p}$ (A/W), convert the incident optical power into a proportional current $R_{\rm p}\textbf{P}_{\rm r}$. Then, the direct current (DC) bias is removed, and the signals are amplified via a transimpedance amplifier of gain $T$ (V/A) to produce an $N_{\rm r} \times 1$ voltage signal vector $\textbf{y}$, which is a scaled, but noisy, version of the transmitted signal $\textbf{s}$. \\
\indent Based on the above, the resulting signal model is described as:
\begin{equation}
\mathbf{y}= \lambda \mathbf {H}\mathbf{x}+\mathbf{n},
\end{equation}
\begin{figure*}[!t]
\centering
\begin{subfigure}[b]{0.55\columnwidth}
\centering
\includegraphics[width=1\linewidth]{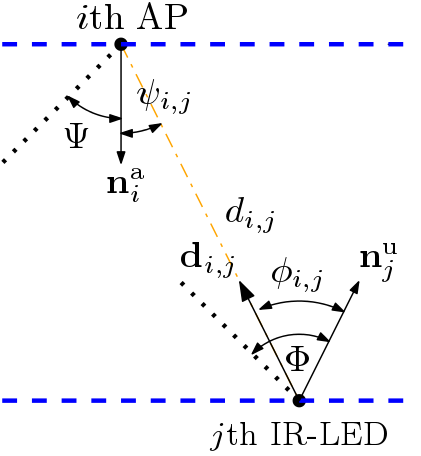}
\caption{LOS}
\label{fig:LOS}
\end{subfigure}
\hspace{1.5cm}
\begin{subfigure}[b]{1\columnwidth}
\centering
\includegraphics[width=1\linewidth]{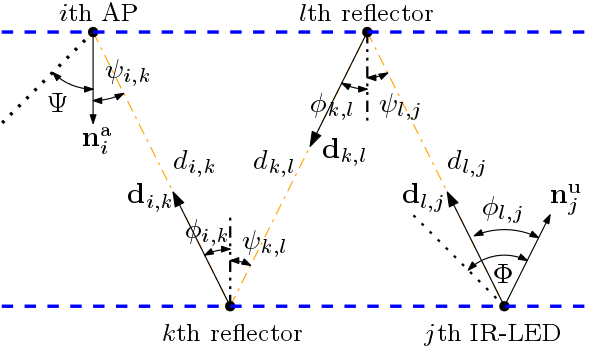}
\caption{NLOS}
\label{fig:NLOS}
\end{subfigure}
\caption{The uplink geometry of optical wireless communications with randomly-orientated user device.}
\label{fig:channelgain}
\end{figure*}
where $\lambda = T R_{\rm P} \eta$ and $\mathbf{n} = [n_1,n_2,...,n_{N_{\rm r}}]^T$ is the $N_{\rm r}\times1$ noise vector at the PDs of the APs, such that for $i \in \llbracket 1,N_{\rm r} \rrbracket$, $n_i$ is the noise experienced at the $i$th PD. The noise here includes all possible noises, such as shot noise and thermal noise and is assumed to be real valued additive white Gaussian $\mathcal{N} \left(\textbf{0}_{N_{\rm r}}, \sigma_{\rm n}^2 \textbf{I}_{N_{\rm r}} \right)$ and independent of the transmitted signal \cite{fath2013performance}. The variance of the noise is equal to $\sigma_{\rm n}^2=N_0B$, where $N_0$ is the single sided power spectral density of noise and $B$ is the bandwidth.
\subsection{Channel Model}
\label{sec:channel_model}
\begin{figure}[t]
		\centering
		\begin{subfigure}[b]{0.45\columnwidth}
			\centering
			\includegraphics[width=\columnwidth,draft=false]{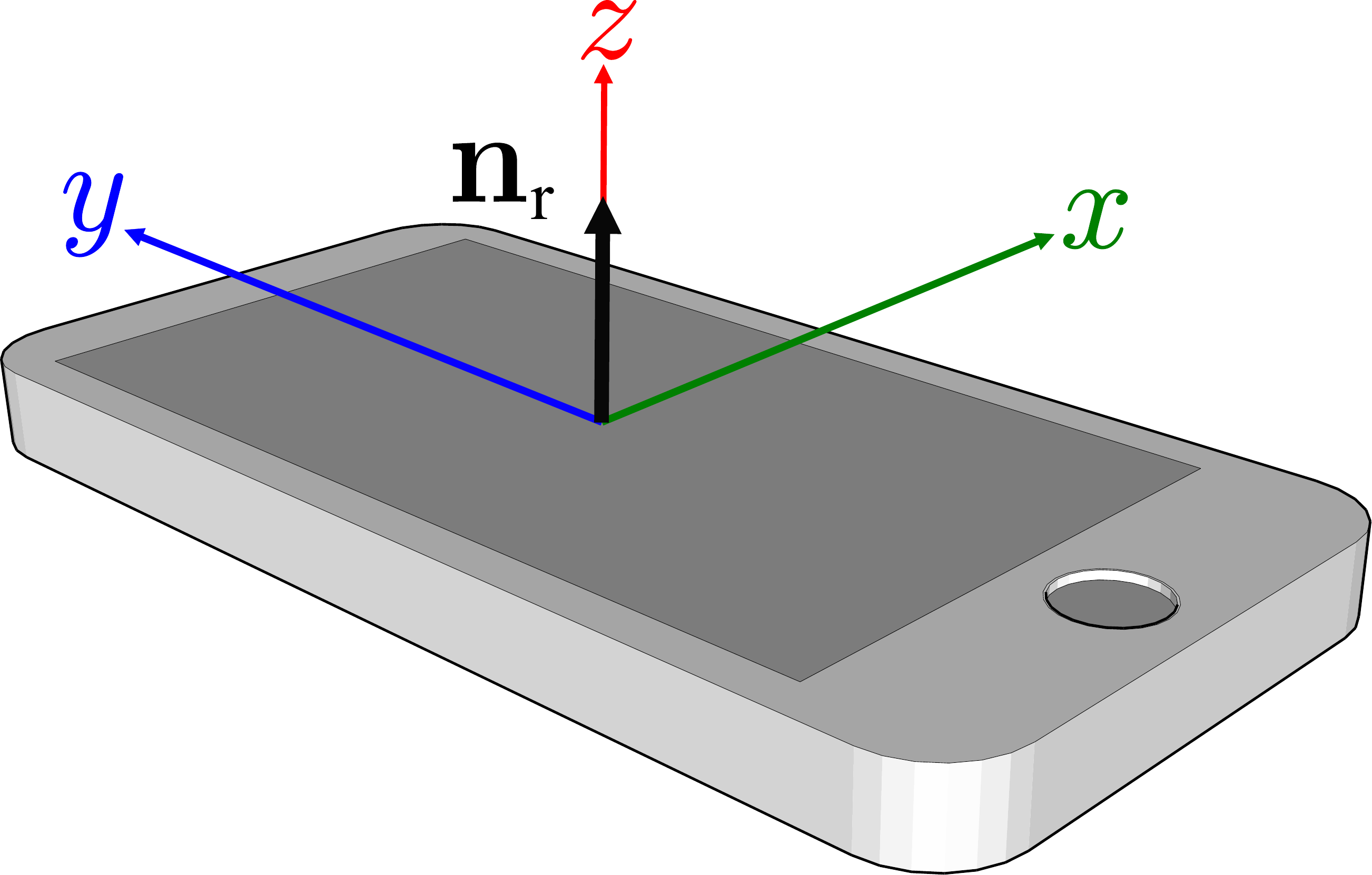}
            \vspace{-0.3cm}
			\caption{}
		\end{subfigure}%
		~
		\begin{subfigure}[b]{0.45\columnwidth}
			\centering
			\includegraphics[width=\columnwidth,draft=false]{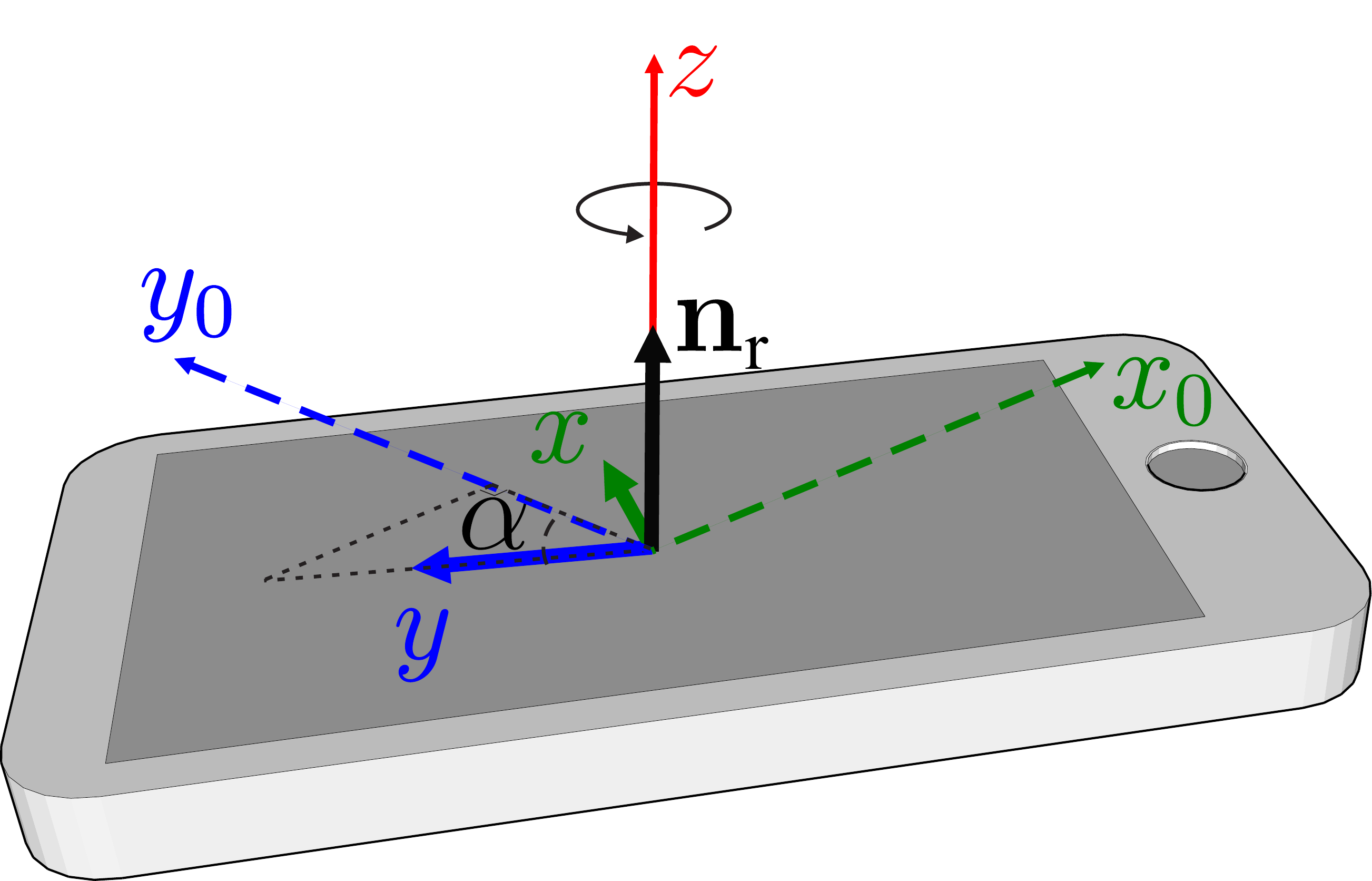}
            \vspace{-0.3cm}
			\caption{}
		\end{subfigure}\\
		\begin{subfigure}[b]{0.45\columnwidth}
			\centering
			\includegraphics[width=\columnwidth,draft=false]{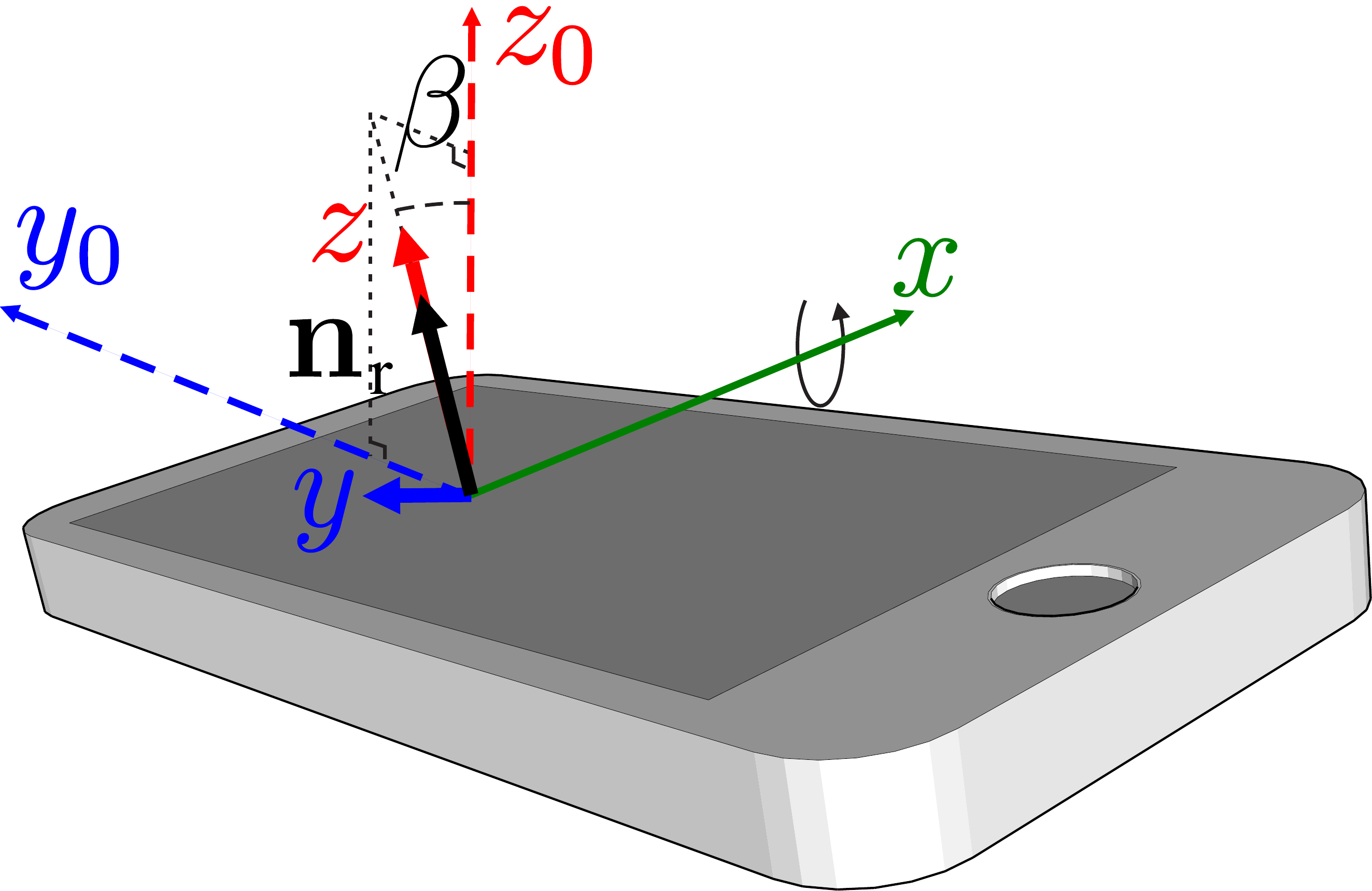}
            \vspace{-0.2cm}
			\caption{}
		\end{subfigure}%
		~
		\begin{subfigure}[b]{0.45\columnwidth}
			\centering
			\includegraphics[width=\columnwidth,draft=false]{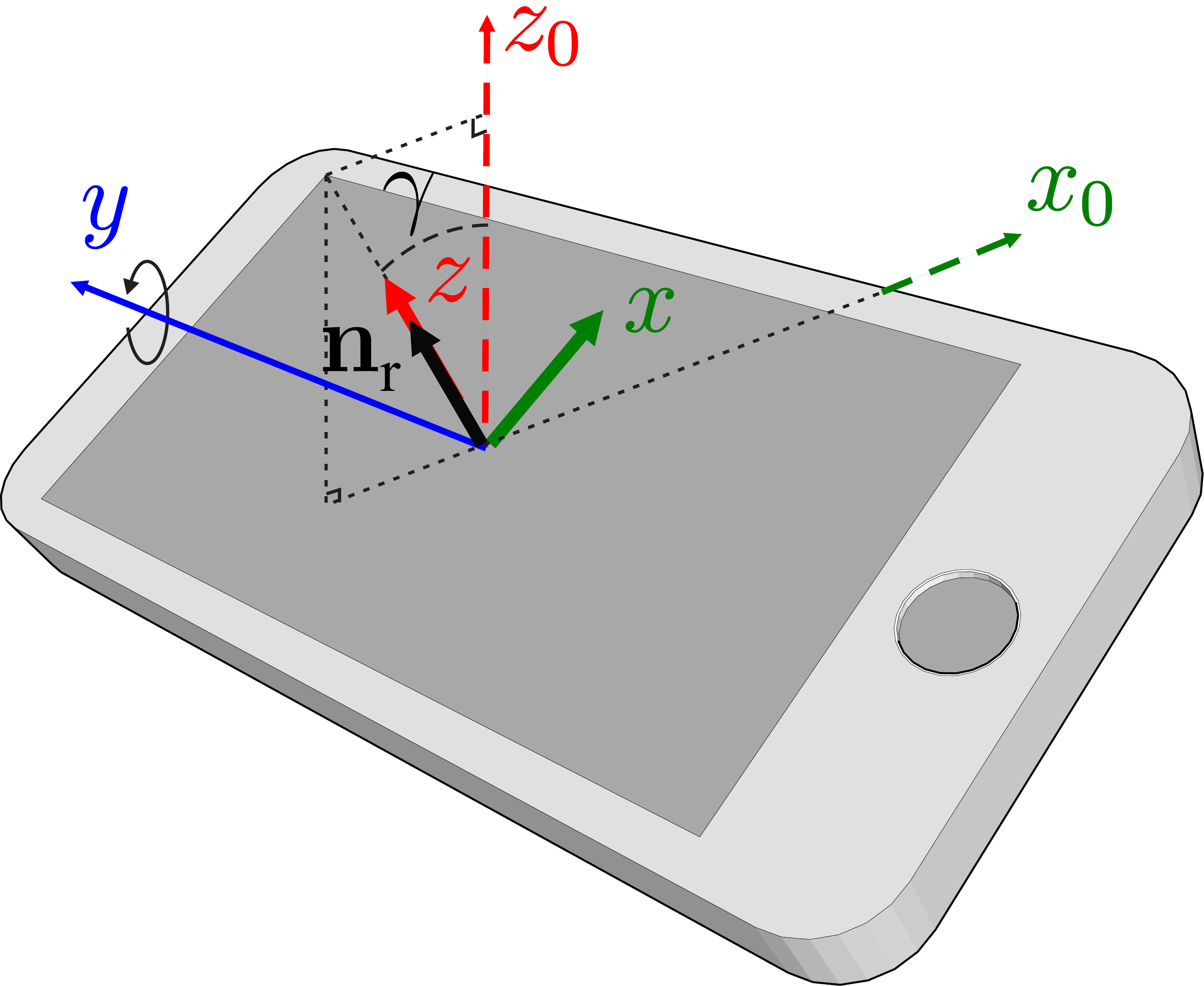}
			\caption{}
		\end{subfigure}
		\caption{Orientation of a mobile device: (a) normal position, (b) yaw rotation with angle $\alpha$, about the $z$-axis (c) pitch rotation with angle $\beta$, about the $x$-axis and (d) roll rotation with angle $\gamma$, about the $y$-axis.}
		\vspace{-0.6cm}
		\label{figori}
	\end{figure}

\indent The channel matrix $\mathbf{H}$ is given by:
\begin{equation}
\mathbf {H}= 
 \begin{pmatrix}
  h_{1,1} &  \cdots & h_{1,N_{\rm t}} \\
  \vdots    & \ddots & \vdots  \\
  h_{N_{\rm r},1} & \cdots & h_{N_{\rm r},N_{\rm t}} 
 \end{pmatrix},
\end{equation}
where, for all $i \in \llbracket 1,N_{\rm r} \rrbracket$ and for all $j \in \llbracket 1,N_{\rm t} \rrbracket$, $h_{i,j}$ is the channel gain of the link between the $j$th IR-LED and the PD of the $i$th AP, which can be expressed as:
\begin{equation}\label{h}
h_{i,j} = h_{i,j}^\mathrm{LOS}+h_{i,j}^\mathrm{NLOS},
\end{equation}
such that $h_{i,j}^\mathrm{LOS}$ and $h_{i,j}^\mathrm{NLOS}$ denote the LOS and the NLOS channel gains, respectively. Each channel gain depends on the positions of the corresponding AP as well as the position and orientation of the corresponding IR-LED, which is explained as follows. \\
\indent For all $i \in \llbracket 1,N_{\rm r} \rrbracket$ and for all $j \in \llbracket 1,N_{\rm t} \rrbracket$, Fig.~\ref{fig:LOS} shows the LOS link geometry between the $j$th IR-LED and the PD of the $i$th AP, where $\textbf{n}_j^{\rm u}$ is the normal vector of the $j$th IR-LED, $\textbf{n}_i^{\rm a}$ is the normal vector of the PD of the $i$th AP, $\phi_{i,j}$ is the angle of radiance, $\Phi$ is the IR-LED field-of-view (FOV), $\psi_{i,j}$ is the incidence angle, $\Psi$ is the FOV of the PD at the AP and $d_{i,j}$ is the distance between the $j$th IR-LED and the PD at the $i$th AP. Based on \cite{kahn1997wireless}, the LOS channel gain $h_{i,j}^\mathrm{LOS}$ is expressed as:
\begin{equation}
\label{eq:LOS}
\begin{split}
&h_{i,j}^\mathrm{LOS}=\\
&\frac{(m+1)A}{2 \pi d_{i,j}^2} \cos^m(\phi_{i,j})\mathrm{rect}\left(\frac{\phi_{i,j}}{\Phi}\right) \cos(\psi_{i,j}) \mathrm{rect}\left(\frac{\psi_{i,j}}{\Psi}\right),
\end{split}
\end{equation}
where $m=-1/\log_2(\cos(\Phi_{1/2}))$ is the Lambertian emission order of the IR-LEDs, such that $\Phi_{1/2}$ is the associated half-power semi-angle, and $A$ is the area of the PD.\\ 
\indent Assuming that the global coordinate system $\left(O,X,Y,Z \right)$ is Cartesian, and for all $i \in \llbracket 1,N_{\rm r} \rrbracket$, the coordinates of the $i$th AP are given by $\left(x_i^{\rm a},y_i^{\rm a},z_i^{\rm a} \right)$, the coordinates of the UE are given by $(x,y,z)$, and for all $j \in \llbracket 1,N_{\rm t} \rrbracket$, the coordinates of the $j$th (IR-LED,PD) pair of the UE are given by $\left(x_j^{\rm u}, y_j^{\rm u}, z_j^{\rm u} \right)$. Basically, the coordinates of the IR-LEDs depend on the geometry of the UE and they are fully known once the 3D position of the UE is known. In other words, for all $j \in \llbracket 1,N_{\rm t} \rrbracket$, $\left(x_j^{\rm u}, y_j^{\rm u}, z_j^{\rm u} \right) = (x+\Delta x_j,y+\Delta y_j,z + \Delta z_j)$, where $\left(\Delta x_j,\Delta y_j,\Delta z_j \right)$ are constant and depend on the design of the UE. Finally, for all $i \in \llbracket 1,N_{\rm r} \rrbracket$ and $j \in \llbracket 1,N_{\rm t} \rrbracket$, we denote by $\textbf{p}_i^{\rm a}$ and $\textbf{p}_j^{\rm u}$ the 3D vectors defined, respectively, as $\textbf{p}_i^{\rm a} \triangleq \left[x_i^{\rm a}, y_i^{\rm a}, z_i^{\rm a}\right]^T$ and $\textbf{p}_j^{\rm u} \triangleq \left[x_j^{\rm u}, y_j^{\rm u}, z_j^{\rm u}\right]^T$, i.e., they contain the position of the $i$th AP and the $j$th IR-LED, respectively. Based on this, for all $i \in \llbracket 1,N_{\rm r} \rrbracket$ and $j \in \llbracket 1,N_{\rm t} \rrbracket$, the cosine of the incidence angle at the $i$th AP and the cosine of the radiance angle from the $j$th IR-LED can be expressed, respectively, as:
\begin{equation}
\cos(\psi_{i,j}) = -\frac{{\textbf{n}_i^{\rm a}}^T \textbf{d}_{i,j}}{||\textbf{d}_{i,j}||}, \quad \text{and} \quad \cos(\phi_{i,j}) =  \frac{{\textbf{n}_j^{\rm u}}^T \textbf{d}_{i,j}}{||\textbf{d}_{i,j}||}.
\end{equation}
\indent Concerning the NLOS components of the channel gain, they can be calculated based on the method described in \cite{NLOSSchulze}. Using the frequency domain instead of the time domain analysis, one is able to consider an infinite number of reflections to have an accurate value of the diffuse link. The environment is segmented into a number of surface elements which reflect the light beams. These surface elements are modeled as Lambertian radiators described by \eqref{eq:LOS} with $m=1$ and FOV of $90^\circ$. Assuming that the entire room can be decomposed into $K$ surface elements, the NLOS channel gain $h_{i,j}^\mathrm{NLOS}$, including an infinite number of reflections between the $j$th IR-LED and the $i$th AP, for all $i \in \llbracket 1,N_{\rm r} \rrbracket$ and $j \in \llbracket 1,N_{\rm t} \rrbracket$,  can be expressed as \cite{NLOSSchulze}:
\begin{equation}
\label{eq:NLOS}
h_{i,j}^\mathrm{NLOS}=\mathbf{r}^\mathrm{T}\mathbf{G}_\zeta(\mathbf{I-EG_\zeta})^{-1}\mathbf{t},
\end{equation} 
where the vectors $\mathbf{t}$ and $\mathbf{r}$ respectively represent the LOS link between the $j$th IR-LEDs and all the surface elements of the room and from all the surface elements of the room to the $i$th AP. The matrix $\mathbf{G}_\zeta={\rm{diag}}(\zeta_1,...,\zeta_K)$ is the reflectivity matrix of all $K$ reflectors; $\mathbf{E}$ is the LOS transfer function of size $K\times K$ for the links between all surface elements, and $\mathbf{I}_K$ is the unity matrix of order $K$. In \eqref{eq:NLOS}, the elements of $\mathbf{E}$, $\mathbf{r}$ and $\mathbf{t}$ are found according to \eqref{eq:LOS} and Fig. \ref{fig:NLOS} between groups of IR-LED, surface elements and PD. In this paper, we assume that the modulation bandwidth is within the 3 dB bandwidth of the optical wireless transmission channel. Therefore, temporal delay between different transmitter-receiver pairs and the temporal dispersion can be neglected. Hence, only the DC channel gain is considered including LOS and NLOS components \cite{mesleh2011optical}. \\ 
\indent The performance of the considered LiFi system depends heavily on the channel matrix $\textbf{H}$. Devices such as laptops are usually placed on a flat surface and the IR-LEDs can be assumed to retain their orientation during each communication session whether upward or not \cite{MDSFeedback}. However, hand-held devices such as smartphones are prone to random changes in orientation due to hand motion. In this study, we focus on these types of devices and incorporate the random orientation in our analysis. The orientation of the LiFi UE is fully characterized in three dimensions through the elemental rotation angles yaw, $\alpha \in [0^\circ,360^\circ)$, pitch, $\beta \in [-180^\circ,180^\circ)$, and roll, $\gamma \in [-90^\circ,90^\circ)$ \cite{MDSHandover}. As shown in Fig.~\ref{figori}, $\alpha$, $\beta$ and $\gamma$ denote the rotations about $Z$-axis, $X$-axis and $Y$-axis, respectively. According to the Euler's rotation theorem, any rotation matrix can be expressed by 
$\mathbf{R}=\mathbf{R}_\alpha \mathbf{R}_\beta \mathbf{R}_\gamma,$
where
\begin{equation}\centering
\begin{split} 
\mathbf{R}_\alpha=\begin{bmatrix}
		\cos \alpha & -\sin \alpha & 0 \\
		\sin \alpha & \cos \alpha & 0 \\
		0 & 0 & 1 
\end{bmatrix},
\mathbf{R}_\beta=\begin{bmatrix}
		1 & 0 & 0 \\
		0 & \cos \beta & -\sin \beta \\
		0 & \sin \beta & \cos \beta 
		\end{bmatrix},\\
\mathbf{R}_\gamma=\begin{bmatrix}
		\cos \gamma & 0 & \sin \gamma \\
		0 & 1 & 0 \\
		-\sin \gamma & 0 & \cos \gamma
				\end{bmatrix}.~~~~~~~~~~~~~~~~~~~~~
				\end{split}
\end{equation}
Hence, for all $j \in \llbracket 1,N_{\rm t} \rrbracket$, the normal vector of the $j$th IR-LED after performing the rotation can be described by $\mathbf{n}_j^{\rm u}=\mathbf{R}\mathbf{n}_j^{\rm u,0}$, where $\mathbf{n}_j^{\rm u,0}$ is the orientation vector of the $j$th IR-LED when the UE is at the standard position, as shown in Fig.~\ref{figori}(a).
\subsection{Objective and RSS Analysis}
\indent The objective of this paper is estimating the 3D position and orientation of a UE communicating in the uplink phase with the $N_{\rm r}$ APs. The UE is randomly located within the indoor environment shown in Fig.~\ref{fig:IndEnv} and its orientation is also varying randomly. In the joint position and orientation estimation process, the UE needs to transmit a reference signal to the APs in a one resource block (time/frequency). Assuming that the DC-biased pulse-amplitude modulation (PAM) with order $M$ is used, the UE broadcasts through its $N_{\rm t}$ IR-LEDs a scalar signal $s$ that is equal to one of the $M$-PAM intensity levels, which are given by:
\begin{equation}
    I_m = \frac{\left(2m - (M+1) \right)}{M+1}I_{\rm DC}, \quad \text{for} \,\, m=1,2,...,M.
\end{equation}
Hence, the transmitted vector of signals is given by $\textbf{s} = s \textbf{1}_{\rm t}$, and for $i \in \llbracket 1,N_{\rm r} \rrbracket$, the received signal at the $i$th AP is given by $y_{i} =  \left(\lambda \sum_{j=1}^{N_{\rm t}} h_{i,j} \right) s + n_{i}$. Consequently, the received SNR at the $i$th AP is given by:
\begin{equation}
    \label{eq:SNR}
    \rho_{i} = \frac{\left(\lambda \sum_{j=1}^{N_{\rm t}} h_{i,j} \right)^2P_{\rm elec}}{\sigma_n^2}, 
\end{equation}
where $P_{\rm elec} = \frac{I_{\rm DC}^2}{3} \frac{M-1}{M+1}$ is the electrical power of the transmitted signal $s$. \\ 
\indent Based on the above, the $N_{\rm r} \times 1$ received SNR vector at the APs, defined as $\boldsymbol{\rho} \triangleq \left[\rho_1, \rho_2,...,\rho_{N_{\rm r}} \right]$ is based on the channel matrix $\textbf{H}$, which in turn depends mainly on six random variables, which are $\left(x,y,z,\alpha,\beta,\gamma\right)$. Precisely, the variables $\left(x,y,z \right)$ model the randomness of the instantaneous position of the UE, whereas the variables $\left(\alpha,\beta,\gamma\right)$ model the randomness of its instantaneous orientation. Such correlation can be exploited in estimating the instantaneous UE position and orientation, which will be elaborated in the following section. 
\section{RSS-Based Fingerprinting for Position and Orientation Estimation}
\subsection{Proposed Approach}
Let us consider the indoor LiFi system shown in Fig.~\ref{fig:IndEnv}, where a UE is randomly located in the room and its orientation is also random. Assuming that the UE is communicating with the APs installed at the ceiling of the room, the objective of this paper is estimating the instantaneous 3D position and orientation of the UE based on the instantaneous received SNRs at the APs. Unlike the several positioning methods reported in the literature, there are no requirements or prior knowledge neither on the UE position and orientation nor on the emitting power of the positioning signals transmitted by the UE. In other words, the variables $\left(x,y,z,\alpha,\beta,\gamma\right)$ along with the UE emitting power $P_{\rm elec}$ are totally unknown without any prior information on them. The adopted estimation technique is fingerprinting and the estimation metric is the received SNR, $\boldsymbol{\rho}$, at the APs. The details of the proposed approach are explained in the following. \\
\indent The proposed joint position and orientation estimation approach is divided into two phases: 1) an offline survey (offline phase) and 2) an online testing (online phase). In the offline survey, the received SNR at the APs for possible 3D positions $\left(x,y,z \right)$ and orientation angles $\left(\alpha,\beta,\gamma \right)$ are collected, processed and recorded into a dataset. Then, based on the obtained measurements-based dataset, optimal learning models that provide the best mappings between the instantaneous received SNR and the 3D position and the orientation are built. In the online testing, the obtained models are tested against the real 3D position and orientation angles of the UE to evaluate the accuracy of the derived models. In the following, we will present first the steps of the offline phase and then we will discuss the deployment of the obtained learning models in the online phase.
\subsection{Dataset Generation} 
\begin{figure}[t]
\centering     
\includegraphics[width=0.5\linewidth]{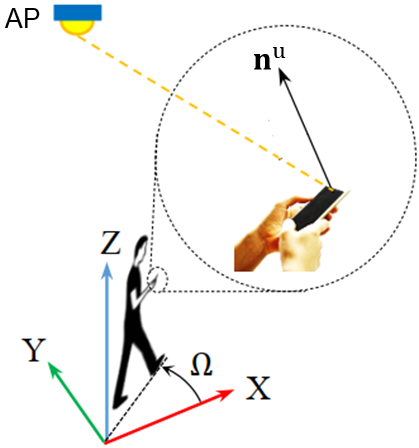}
\caption{User direction.}
\label{fig:user_direction}
\end{figure}

\begin{figure}[t]
\centering     
\includegraphics[width=1\linewidth]{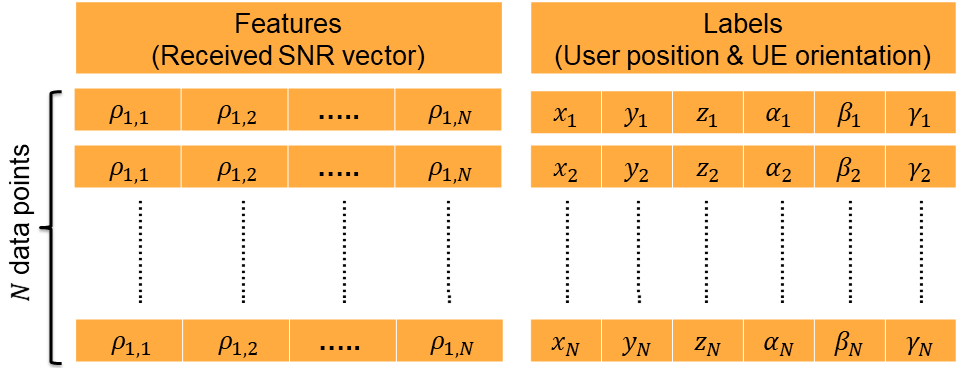}
\caption{Dataset architecture.}
\label{fig:data_set}
\end{figure}

\indent Each UE is assumed to be stationary within the indoor environment. In this case, and as shown in \cite{arfaoui2020measurements}, the user is uniformly located within the indoor environment, and therefore, the PDFs of the UE 3D position are given by: 
\begin{subequations}
\label{eq:position_statistics}
\begin{align}
    &f_x(x) = \frac{1}{L}\mathcal{U}_{[-\frac{L}{2},\frac{L}{2}]}\left( x \right),  \\
    &f_y(y) = \frac{1}{W}\mathcal{U}_{[-\frac{W}{2},\frac{W}{2}]}\left( y \right),  \\
    &f_z(z) = \frac{1}{H_{\rm device}}\mathcal{U}_{[0,H_{\rm device}]}\left( z \right),  
\end{align}
\end{subequations}
where $0 \leq H_{\rm device} \leq H$ is the maximum height of any UE within the indoor environment. \\ 
\indent On the other hand, a set of experiments were conducted in \cite{soltani2018modeling,Zhihong_VTCfall_2018,purwita2018WCNC}, aiming to derive some accurate measurements-based statistical models for the rotation angles $\left(\alpha,\beta,\gamma \right)$. For collecting the measurements, $40$ participants were asked to take part in the experiment while they were working with their cellphones. The application ``physics toolbox sensor suite" \cite{androidapp} was used to record the orientation data of yaw $\alpha$, pitch $\beta$ and roll $\gamma$ while users were doing normal activities like browsing or watching a video stream. Measurements were recorded for static and mobile users (sitting and walking activities, respectively). More details about the data measurement can be found in \cite{soltani2018modeling,Zhihong_VTCfall_2018,purwita2018WCNC}.\\ 
\indent Based on the results of \cite{soltani2018modeling,Zhihong_VTCfall_2018,purwita2018WCNC}, the rotation angles $\alpha$, $\beta$ and $\gamma$, follow each a truncated Laplace distribution with mean and standard deviation $\left(\mu_\alpha, \sigma_\alpha \right)$, $\left(\mu_\beta, \sigma_\beta \right)$ and $\left(\mu_\beta, \sigma_\beta \right)$, respectively, which are presented in Table~\ref{distfit}, where $\Omega$ denotes the movement direction. Precisely, $\Omega$ denotes the facing or movement direction of a user while sitting or walking, which is measured from the East direction in the Earth coordinate system as shown in Fig.~\ref{fig:user_direction}. From a statistical point of view, it was also shown in \cite{soltani2018modeling} that the movement direction angle $\Omega$ follows a uniform distribution within $[0^\circ,360^\circ]$.\\
\begin{table}[t!]
	\caption{\small Statistics of orientation measurement.}
	\label{distfit}
	\centering
	\begin{tabular}{c c c c}
		\hline
		\hline
		& $\alpha$& $\beta$ & $\gamma$ \\
		\hline
		Mean & $\Omega$-90 &40.78& -0.84 \\
		Standard deviation& 3.67& 2.39 &2.21\\
		\hline
		\hline
	\end{tabular}
\end{table}
\indent Based on the above, assuming the target dataset contains $N$ data points, the procedure of generating the $n$th measurement-based data point, for $n \in \llbracket 1,N \rrbracket$,  are detailed as follows.
\begin{enumerate}
    \item A sample of 3D position $(x,y,z)$ is generated using the statistics in \eqref{eq:position_statistics}. 
    \item A sample of movement direction angle $\Omega$ is generated uniformly from $[0^\circ,360^\circ]$.
    \item The three orientation angles $\left(\alpha,\beta,\gamma \right)$ are generated using the truncated Laplace distribution and the statistics specifications in Table~\ref{distfit}.
    \item The resulting channel matrix $\textbf{H}$ is then calculated as explained in subsection \ref{sec:channel_model}.
    \item A random electrical emission power $P_{\rm elec}$ is generated uniformly from $[0,P_{\rm elec}^{\max}]$, where $P_{\rm elec}^{\max}$ is the highest possible electrical emission power from the UE.
    \item The corresponding SNR vector $\boldsymbol{\rho}$ is calculated as shown in \eqref{eq:SNR}.
    \item Finally, the resulting SNR vector $\boldsymbol{\rho}$ is stored into the dataset as a feature vector and the corresponding 3D-position and orientation angles $(x,y,z,\alpha,\beta,\gamma)$ are stored as a label vector as shown in Fig.~\ref{fig:data_set}.
\end{enumerate}
\indent Once the data set is ready, the goal now is how to obtain ``good" mappings between the feature vector that contains the received SNR and the label vector that contains the 3D position and the orientation angles of the UE using the obtained dataset. For such a goal, several learning methods can be applied such as KNN \cite{tran2020high}, SVM and ELM \cite{chen2019indoor,wang2019machine}. In this paper, we will employ, and for the first time, deep ANNs, as it will be presented in the following subsection.
\subsection{Learning Models: Deep ANNs}
\begin{figure}[t]
\centering     
\includegraphics[width=0.9\linewidth]{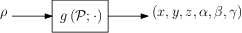}
\caption{Estimation mapping between the received SNR and the 3D position and orientation angles of the UE.}
\label{fig:mapping}
\end{figure}
\begin{figure}[t]
\centering     
\includegraphics[width=1\linewidth]{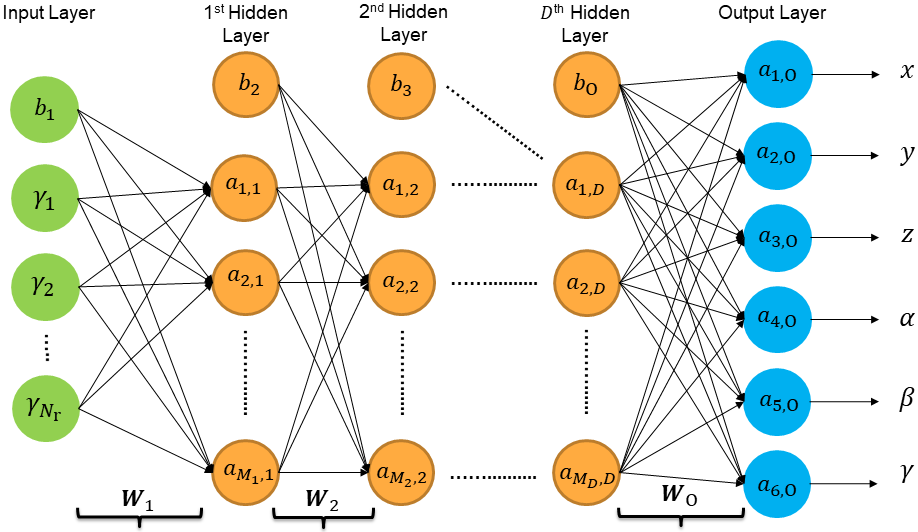}
\caption{ANN architecture for joint 3D position and orientation estimation.}
\label{fig:MLP}
\end{figure}

\indent ANNs have been widely applied in various fields to overcome the problem of complex and nonlinear mappings. Recently, different kinds of ANNs have been applied in a wide range of applications in wireless communication, especially in the physical layer, such as modulation recognition, channel modeling, signal processing and data decoding \cite{ali20206g}. In our context, and as it can be seen in Fig.~\ref{fig:mapping}, our objective is to find a parametric mapping $g(\mathcal{P};\cdot)$, where $\mathcal{P}$ represents a set of parameters, that can link between the instantaneous received SNR vector $\boldsymbol{\rho}$ in one hand and the 3D position and the orientation angles of the UE $\left(x,y,z,\alpha,\beta,\gamma\right)$ in the other hand. Using ANNs, $g(\mathcal{P};\cdot)$ is indeed a neural network and such mapping can be obtained by determining the optimal set of parameters $\mathcal{P}^*$ that produce the best mapping with respect to a given estimation error metric, i.e.,
\begin{equation}
    (x,y,z,\alpha,\beta,\gamma) = g\left(\mathcal{P}^*,{\rm \boldsymbol{\rho} } \right).
\end{equation}
For this mission, two different models of ANNs are considered in this paper, which are the multilayer perceptron (MLP) and the convolutional neural network (CNN). In the following, we present the architectures of MLP and CNN models. \\
\indent An ANN is a series of layers, where each layer is composed of multiple artificial neurons and their connections. Specifically, as shown in Fig.~\ref{fig:MLP}, an ANN is composed of an input layer, $D$ hidden layers and an output layer, where $D$ denotes the depth of the neural network. First, at the input layer, the SNR feature vector $\boldsymbol{\rho}$ with a bias $\textbf{b}_1$ is fed into the neural network. Second, for $d=1,2,...,D$, the $d$th hidden layer consists of $M_d$ artificial neurons and their connection. Each artificial neuron has the ability to calculate a mathematical operation of its inputs and then applies an activation function to obtain a signal that will be forwarded to the next layer. Finally, the output layer consists of $6$ artificial neurons, where each neuron is responsible for estimating one parameter in $(x,y,z,\alpha,\beta,\gamma)$. As shown in Fig.~\ref{fig:MLP}, the propagation rules within the hidden layers are expressed as follows. For $j \in \llbracket 1,D \rrbracket$ and $i \in \llbracket 1,M_j \rrbracket$, the output of the $i$th neuron in the $j$th hidden layer is expressed as
\begin{equation}
    v_{i,j} = a_{i,j} \left[{\rm t} \left(\textbf{u}_{j-1}, \textbf{w}_{i,j}, b_{i,j}\right) \right],
\end{equation}
where $a_{i,j}[\cdot]$, $\textbf{u}_{j-1}$ and $\textbf{w}_{i,j}$ denote the activation function, the input and the weights vector of the $i$th neuron of the $j$th hidden layer, respectively, ${\rm t}(\cdot,\cdot,\cdot)$ is a linear transformation that depends on the type of the ANN and $b_{i,j}$ is a scalar bias. Assuming that the output layer is referred to as the $(D+1)$th layer, note that for $j \in \llbracket 0,D \rrbracket$, the input vector $\textbf{u}_j$ of the $(j+1)$th layer is exactly the output vector $\textbf{v}_j$ of $j$th layer, i.e., 
\begin{equation}
    \textbf{u}_{j} = \textbf{v}_{j} = \left[v_{1,j},v_{2,j},...,v_{M_{j},j} \right]^T,
\end{equation}
with the convention $\textbf{u}_0 = \textbf{v}_0 = \boldsymbol{\rho}$, which is the SNR feature vector. On the other hand, the propagation rules in the output layer are given by
\begin{equation}
    \begin{aligned}
    x &= a_{1,{\rm o}} \left[ {\rm t} \left(\textbf{u}_{D}, \textbf{w}_{1,{\rm o}}, b_{i,\rm O} \right)\right], \quad &\alpha = a_{4,{\rm o}} \left[ {\rm t} \left(\textbf{u}_{D}, \textbf{w}_{4,{\rm o}}, b_{i,\rm O} \right)\right], \\ 
    y &= a_{2,{\rm o}} \left[ {\rm t} \left(\textbf{u}_{D}, \textbf{w}_{2,{\rm o}}, b_{i,\rm O} \right)\right], \quad &\beta = a_{5,{\rm o}} \left[ {\rm t} \left(\textbf{u}_{D}, \textbf{w}_{5,{\rm o}}, b_{i,\rm O} \right)\right],\\ 
    z &= a_{3,{\rm o}} \left[ {\rm t} \left(\textbf{u}_{D}, \textbf{w}_{3,{\rm o}}, b_{i,\rm O} \right)\right], \quad &\gamma = a_{6,{\rm o}} \left[ {\rm t} \left(\textbf{u}_{D}, \textbf{w}_{6,{\rm o}}, b_{i,\rm O} \right)\right],
    \end{aligned}
\end{equation}
where for $k \in \llbracket 1,6 \rrbracket$, $a_{k,{\rm O}}(\cdot,\cdot,\cdot)$ and $\textbf{w}_{k,{\rm o}}$, denote the activation function and the weights vector of the $k$th neuron of the output layer and $b_{i,\rm O}$ is a scalar bias. \\ 
\indent The set of parameters $P$ that defines the ANN, is given by $\mathcal{P} = \mathcal{W} \cup \mathcal{B}$, where $\mathcal{W}= \left\{\textbf{W}_{\rm O},\textbf{W}_j|j \in \llbracket 1,D \rrbracket \right\}$, such that $\textbf{W}_{\rm O} = \left[\textbf{w}_{j,{\rm O}},\textbf{w}_{2,{\rm O}},...,\textbf{w}_{M_{\rm O},{\rm O}} \right]$, and for all $j \in \llbracket 1,D \rrbracket$, $\textbf{W}_j = \left[\textbf{w}_{1,j},\textbf{w}_{2,j},...,\textbf{w}_{M_j,j} \right]$, and $\mathcal{B}= \left\{\textbf{b}_{\rm O},\textbf{b}_j|j \in \llbracket 1,D \rrbracket \right\}$, such that $\textbf{b}_{\rm O} = \left[\textbf{b}_{j,{\rm O}},\textbf{b}_{2,{\rm O}},...,\textbf{b}_{M_{\rm O},{\rm O}} \right]^T$, and for all $j \in \llbracket 1,D \rrbracket$, $\textbf{b}_j = \left[\textbf{b}_{1,j},\textbf{b}_{2,j},...,\textbf{b}_{M_j,j} \right]^T$. As was mentioned above, the linear transformation ${\rm t}(\cdot,\cdot,\cdot)$ depends on the type of the ANN used. In our approach, we distinguish between two main linear transformations, which are the weighted sum and the 1D convolution. The ANN that employs the weighted sum as a linear transformation is the multilayer perceptron (MLP). In this case, for $j \in \llbracket 1,D \rrbracket$, $i \in \llbracket 1,M_j \rrbracket$ and $k \in \llbracket 1,3 \rrbracket$, the linear transformation ${\rm t}(\cdot,\cdot,\cdot)$ is defined as
\begin{equation}
    \left\{
    \begin{aligned}
    &{\rm t} \left(\textbf{u}_{j-1}, \textbf{w}_{i,j}, b_{i,j}\right) = \textbf{w}_{i,j}^T\textbf{u}_{j-1}+b_{i,j}, \\ 
    &{\rm t} \left(\textbf{u}_{D}, \textbf{w}_{k,{\rm o}}, b_{i,\rm O} \right) = \textbf{w}_{k,{\rm o}}^T\textbf{u}_{D}+b_{i,\rm O}.
    \end{aligned}
    \right.
\end{equation}
On the other hand, the ANN that employs the 1D convolution as a linear transformation is the convolutional neural network (CNN). In this case, for $j \in \llbracket 1,D \rrbracket$, $i \in \llbracket 1,M_j \rrbracket$ and $k \in \llbracket 1,3 \rrbracket$, the linear transformation ${\rm t}(\cdot,\cdot,\cdot)$ is defined as
\begin{equation}
    \left\{
    \begin{aligned}
    &{\rm t} \left(\textbf{u}_{j-1}, \textbf{w}_{i,j}, b_{j}\right) = \textbf{w}_{i,j} \circledast \textbf{u}_{j-1}+b_{i,j}, \\ 
    &{\rm t} \left(\textbf{u}_{D}, \textbf{w}_{k,{\rm o}}, b_{\rm O} \right) = \textbf{w}_{k,{\rm o}} \circledast \textbf{u}_{D}+b_{i,\rm O}, 
    \end{aligned}
    \right.
\end{equation}
where $\circledast$ denotes the convolution operator. \\
\indent The activation function, also known as the threshold function or the transfer function, is a scalar-to-scalar function that determines the output of each neuron in a neural network. The function is attached to each neuron in the network, and determines whether it should be activated or not, based on whether each neuron’s input is relevant for the model’s estimation or not. Some of the most commonly used activation functions for solving non-linear problems include linear function, rectified linear unit (Relu) function, sigmoid function, Hyperbolic tangent, etc \cite{sonoda2017neural}. \\ 
\indent At this stage, the architecture of the ANN model, either MLP or CNN, is set up. The next step is how the ANN model should be trained in a way that provides the best estimation accuracy for the 3D position and the orientation angles of the UE. This is detailed in the following subsection.
\subsection{Models Training}
Once the ANN model is selected, i.e., either MLP or CNN, the goal now is how to obtain the optimal sets of weights $\mathcal{P}^*$ that can map between the instantaneous received SNR vector $\boldsymbol{\rho}$ in one hand and the 3D position and the orientation angles of the UE $\left(x,y,z,\alpha,\beta,\gamma\right)$. This can be obtained by training the selected model as explained in the following. In the estimation (or regression) problem in hands, obtaining the optimal sets of weights is performed by minimizing a certain loss function. In a typical regression problem, several loss functions can be considered, such as the mean-square-error (MSE) or the mean-absolute-error (MAE) \cite{grover2019regression}. In our analysis, we consider the MSE loss, also known as the L$_2$ loss. Hence, obtaining the optimal sets of weights $\mathcal{P}$ can be obtained as
\begin{equation}
    \label{optprob}
    \begin{split}
    \mathcal{P}^* &= \underset{\mathcal{P}}{\rm argmin} \,\, {\rm L}_2 (\mathcal{P}), \\ 
    &=  \underset{\mathcal{P}}{\rm argmin} \,\, \frac{1}{N_{\rm train}} \sum_{l=1}^{N_{\rm train}} ||\textbf{P}_l - \widehat{\textbf{P}}_l\left(\mathcal{P},\boldsymbol{\rho}_l\right)||_2^2,
    \end{split}
\end{equation}
where $N_{\rm train} \in \llbracket 1,N \rrbracket$ is the number of data points used for training the models, and for $l \in \llbracket 1,N_{\rm train} \rrbracket$, $P_l = [x_l,y_l,z_l,\alpha_l,\beta_l,\gamma_l]^T$ and $\widehat{\textbf{P}}_l = \left[\widehat{x}_l,\widehat{y}_l,\widehat{z}_l,\widehat{\alpha}_l,\widehat{\beta}_l,\widehat{\gamma}_l\right]^T$ are the true and estimated label vectors associated to the $l$th feature vector $\rho_l$ of the dataset, respectively, such that $\widehat{\textbf{P}}_l$ is obtained from the selected ANN with respect to the set of parameters $\mathcal{P}$.  \\
\indent Solving the optimization problem in \eqref{optprob} can be performed using the gradient descent algorithm. In fact, gradient descent can be used to minimize the loss function L$_2$ by iteratively moving in the direction of steepest descent as defined by the negative of the gradient \cite{ruder2016overview}. A variety of the gradient descent method is the stochastic gradient descent (SGD), which updates the weight parameters after evaluation of the loss function L$_2$ after each sample. That is, rather than summing up the loss function results for all the samples then taking the mean, SGD updates the weights after every training sample is analysed \cite{ruder2016overview}. Morevoer, several adaptive varieties of the SGD have been proposed in the literature of learning neural networks aiming at either increasing the convergence speed and/or the convergence accuracy, such as Adagrad, Adadelta, RMSprop and Adam, with Adam being the de facto standard in deep learning \cite{ruder2016overview}. 
\subsection{Models Testing}
\indent Once the ANN models are trained and the optimal parameters $\mathcal{P}^*$ are obtained, the models will be deployed in the online phase. Consequently, the performance of the ANN models will be tested over new received SNR vectors (unseen data). In this case, the performance of each ANN model can be evaluated in terms of the following performance metrics: 
\begin{enumerate}
    \item The average estimation error: it represents the average gap between the true label vectors and the estimated label vectors. 
    \item Precision: it represents the estimation error that is higher than $90\%$ of the possible estimation errors
    \item The computational time: it measures the average time needed to estimate the label vector of a given feature vector during one estimation session in the online phase.
\end{enumerate}
In the next section, we evaluate the performance of the proposed ANN models with respect to the above performance metrics.
\section{Simulation Results}
\subsection{Simulations Parameters}
\indent In this paper, we consider a typical indoor environment with dimensions $L \times W \times H$ = $5\times 5 \times 3$ m$^3$ \cite{mohammad2018optical}. Unless otherwise stated, the indoor environment is equipped with $N_{\rm r} = 16$ APs which are arranged on the vertexes of a square lattice over the ceiling of the room, where each AP is oriented vertically downward. In addition, a LiFi UE, that is equipped with $N_{\rm t} = 1$ IR-LED, is randomly located within the room and its UE may have a random orientation. The UE is a typical smartphone with dimensions $14 \times 7 \times 1$ cm$^3$. The IR-LED is placed at screen of the smartphone, exactly at $6$ cm above the center. The parameters used throughout the paper are shown in Table \ref{T2}. 
\begin{table}[t]
\caption{Simulation Parameters}
\centering
\renewcommand{\arraystretch}{1} 
\setlength{\tabcolsep}{0.18cm} 
\begin{tabular}{| c | c | c |}
  \hline 
  Parameter & Symbol & Value \\
  \hline
  Room dimension & $L\times W\times H$ & $5$m$\times 5$m$\times 3$m \\ 
  \hline
  LED half-power semiangle & $\Phi_{1/2}$ & $60^\circ$\\ 
  \hline
  PD responsivity & $R_p$ & $0.6$ A/W \\ 
  \hline
  PD geometric area & $A_g$ & $1$ cm$^2$ \\ 
  \hline 
  Optical concentrator refractive index & $n_c$ & 1 \\
  \hline 
  Maximum UE's height & $H_{\rm device}$ & $1.5$m \\ 
  \hline
  Maximum UE's power & $P_{\rm elec}^{\max}$ & $0.01$ W \\
  \hline 
  Reflection coefficient of the walls & $\zeta$ & $0.7$ \\
  \hline
  Field of view of the IR-LEDs & $\Phi$ & $90^\circ$ \\ 
  \hline 
  Field of view of the PDs & $\Psi$ & $90^\circ$ \\ 
  \hline
  System Bandwidth & $B$ & $10$ MHz \\ 
  \hline 
  Noise power spectral density & $N_0$ & $10^{-21}$ W/Hz \\ 
  \hline
\end{tabular} 
\label{T2}
\end{table}
\begin{table}[t]
\caption{ANNs Specifications}
\centering
\renewcommand{\arraystretch}{1} 
\setlength{\tabcolsep}{0.18cm} 
\begin{tabular}{| c | c |}
  \hline
  \multirow{2}{*}{Dataset size} & First dataset: $N = 10^5$ \\ 
  & Second dataset: $N = 10^6$ \\
  \hline 
  Depth of the ANN $D$ & 4 \\ 
  \hline 
  \multirow{2}{*}{Number or neuron per hidden layer $M_j$} & MLP: 256 \\
  & CNN: 64 \\ 
  \hline 
  Kernel size for CNN neuron & 16 \\
  \hline 
  \multirow{2}{*}{Total number of trainable parameters} & MLP: $207,36$ \\ 
  & CNN: $205,062$ \\
  \hline 
  (Train, test) partition & $(0.9,0.1)\times N$ \\
  \hline
  Optimizer & Adam \\ 
  \hline
\end{tabular} 
\label{T3}
\end{table}
\subsection{ANNs Specifications}
\begin{figure}[t]
		\centering
		\begin{subfigure}[b]{0.3\columnwidth}
			\centering
			\includegraphics[width=\columnwidth,draft=false]{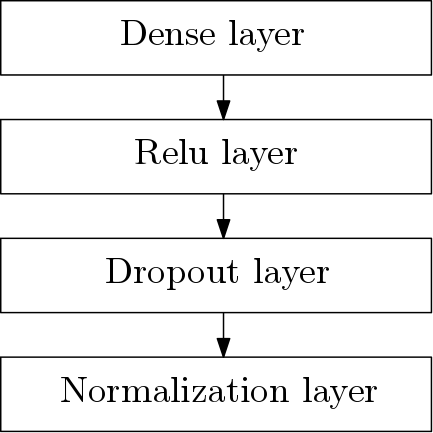}
			\caption{}
		\end{subfigure}
		\begin{subfigure}[b]{0.3\columnwidth}
			\centering
			\includegraphics[width=\columnwidth,draft=false]{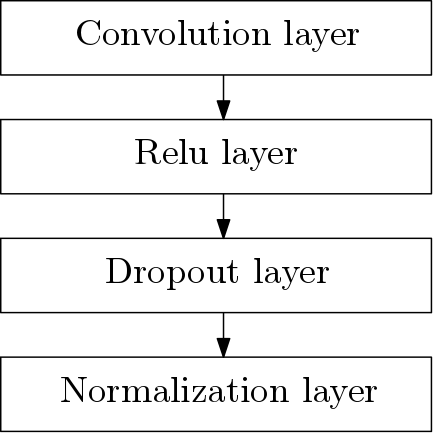}
			\caption{}
		\end{subfigure}
		\begin{subfigure}[b]{0.3\columnwidth}
			\centering
			\includegraphics[width=\columnwidth,draft=false]{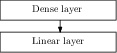}
			\caption{}
		\end{subfigure}
		\caption{Neuron architecture: (a) in each hidden layer in the MLP, (b) in each hidden layer in the CNN and (c) in the output layer.}
		\label{fig:layers}
\end{figure}

\indent The architecture of the two ANNs models MLP and CNN are shown in Table \ref{T3}. Each ANN consists of an input layer, an output layer and $D = 4$ hidden layers, where each hidden layer is composed of $M_j = 256$ neurons for the MLP and $M_j = 64$ filters for the CNN. In addition, the kernel size of each convolution neuron is composed of $16$ parameters. Two distinct datasets are used in training the models, where the first has a size of $N = 10^5$ data points and the second has a size of $N = 10^6$ data points. The required codes for generating the datasets are given in \cite{codes}. \\
\indent In our proposed ANNs, the structure of each neuron in each hidden layer for the MLP and the CNN models are presented in Fig.~\ref{fig:layers}. For the MLP, and as shown in Fig.~\ref{fig:layers}(a), each neuron consists of a dense layer, a Relu layer, a dropout layer and a normalization layer. The dense layer is a fully connected linear layer in which every input is connected to every output by a weight. Then, the Relu layer applies the Relu activation function to the resulting output from the dense layer. After this, at each training stage, individual neurons are dropped out of the ANN with a certain probability, so that the network is reducing. This layer is fundamental in order to prevent overfitting of the ANNs \cite{srivastava2014dropout}. In fact, a fully connected layer occupies most of the parameters, and hence, neurons develop co-dependency amongst each other during training which curbs the individual power of each neuron leading to overfitting of the training data. Finally, The normalization layer scales the input so that the output has near to a zero mean and unit standard deviation, to allow a faster and a more resilient training. For the CNN, and as shown in Fig.~\ref{fig:layers}(b), a similar neuron architecture can be observed but with replacing the dense layer with a convolution layer, which operates the convolution between the input of the neuron and its kernel. Concerning the neurons architecture at the output layer, and as shown in Fig.~\ref{fig:layers}(c), each one consists of a dense layer and a linear activation layer that establishes a link with each label at the output. The design of the ANNs is performed using the programming environment Python 3 and the Keras library developed by Google's TensorFlow team in 2017 \cite{keras}. The codes for designing, training and testing the proposed ANN models are provided in \cite{codes}.
\begin{table*}[t]
\caption{Performance comparison of the proposed MLP and CNN models versus the KNN technique.}
\centering
\renewcommand{\arraystretch}{1} 
\setlength{\tabcolsep}{0.18cm} 
\begin{tabular}{| c | c | c | c | c | c | c | c | c |}
  \cline{4-9} 
  \multicolumn{3}{c}{} & \multicolumn{2}{|c|}{CNN} & \multicolumn{2}{c|}{MLP} & \multicolumn{2}{c|}{KNN} \\ 
  \cline{4-9} 
    \multicolumn{3}{c|}{} & $N=10^5$ & $N=10^6$ & $N=10^5$ & $N=10^6$ & $N=10^5$ & $N=10^6$ \\
    \hline
    \multirow{4}{*}{Position} & \multirow{2}{*}{Average Error [cm]} & LOS & \textbf{21.83} & \textbf{16.49}  & 29.73& 21.93 & 34.71 & 22.16\\
    \cline{3-9} 
    & & LOS + NLOS & \textbf{14.55} & \textbf{10.53} & 15.05 & 13.04 & 27.30 & 17.34\\ 
    \cline{2-9} 
     & \multirow{2}{*}{Precision [cm]} & LOS & \textbf{38.7} & \textbf{29.8} & 49.9 & 37.8 & 61.7 & 40 \\
    \cline{3-9} 
    & & LOS + NLOS & \textbf{23.9} & \textbf{17.15} & 25.1 & 21.4 & 46.5 & 29.5 \\ 
    \hline
    \multirow{4}{*}{Yaw angle $\alpha$} & \multirow{2}{*}{Average Error [Deg]} & LOS & \textbf{15.15} & \textbf{11.9} & 16.40 & 11.68 & 19.05 & 13.30 \\
    \cline{3-9} 
    & & LOS + NLOS & \textbf{12.28} & \textbf{9.07} & 12.56 & 10.09 & 21.37 & 14.49 \\ 
    \cline{2-9} 
     & \multirow{2}{*}{Precision [Deg]} & LOS & \textbf{25} & \textbf{16.9} & 25.5 & 16.67 & 36.1 & 17.2 \\
    \cline{3-9} 
    & & LOS + NLOS & \textbf{18.5} & \textbf{12.5} & 18.9 & 15.5 & 55 & 17.7 \\ 
    \hline
    \multirow{2}{*}{Pitch angle $\beta$} & \multirow{2}{*}{Average Error [Deg]} & LOS & \textbf{1.55} & \textbf{1.42} & 1.61 & 1.5 & 1.88 & 1.7 \\
    \cline{3-9} 
    & & LOS + NLOS & \textbf{1.35} & \textbf{0.96} & 1.40 & 0.98 & 1.88 & 1.8 \\ 
    \cline{2-9} 
     \multirow{2}{*}{$\&$ roll angle $\gamma$} & \multirow{2}{*}{Precision [Deg]} & LOS & \textbf{3.47} & \textbf{3.19} & 3.57 & 3.4 & 4.1 & 3.86 \\
    \cline{3-9} 
    & & LOS + NLOS & \textbf{2.96} & \textbf{2.135} & 3.09 & 2.17 & 4.07 & 3.925 \\
    \hline 
\end{tabular} 
\label{T4}
\end{table*}
\subsection{Learning and Estimation Performance Evaluation}
\begin{figure}[t]
		\centering
		\begin{subfigure}[b]{0.5\columnwidth}
			\centering
			\includegraphics[width=\columnwidth,draft=false]{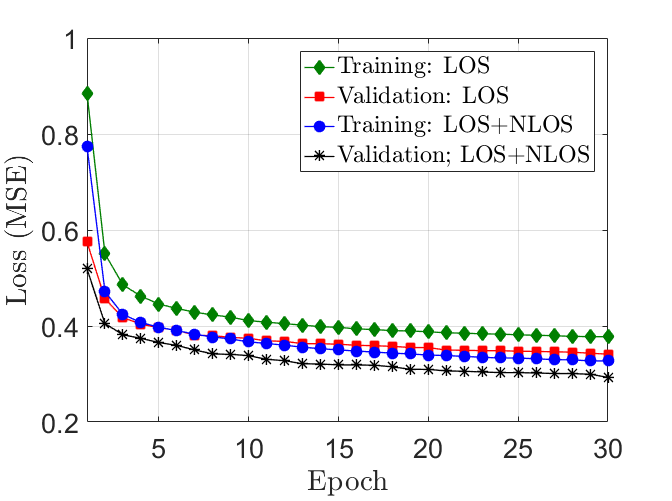}
			\caption{CNN, $N = 10^5$}
		\end{subfigure}%
		~
		\begin{subfigure}[b]{0.5\columnwidth}
			\centering
			\includegraphics[width=\columnwidth,draft=false]{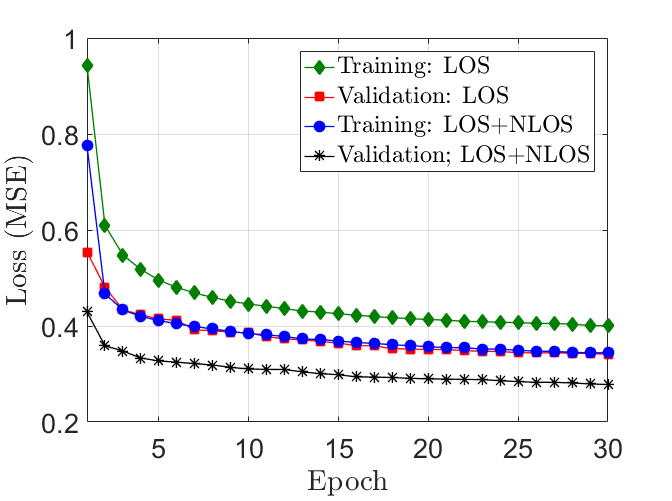}
			\caption{MLP, $N = 10^5$}
		\end{subfigure}\\
		\begin{subfigure}[b]{0.5\columnwidth}
			\centering
			\includegraphics[width=\columnwidth,draft=false]{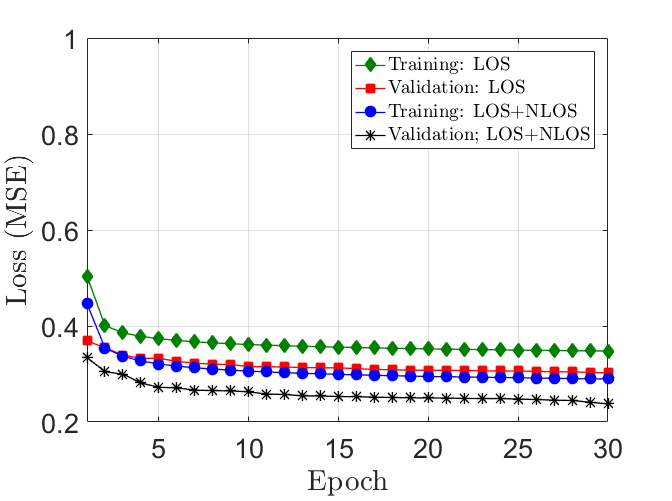}
			\caption{CNN, $N = 10^6$}
		\end{subfigure}%
		~
		\begin{subfigure}[b]{0.5\columnwidth}
			\centering
			\includegraphics[width=\columnwidth,draft=false]{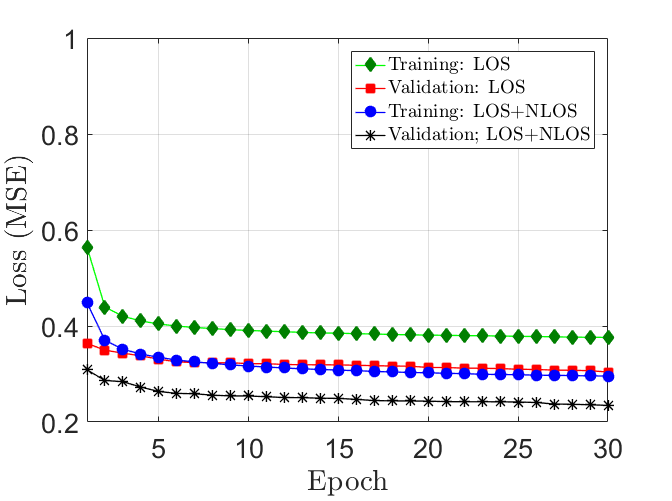}
			\caption{MLP, $N = 10^6$}
		\end{subfigure}
		\caption{Training and validation losses of CNN and MLP models versus the epoch index.}
		\label{fig:Loss}
	\end{figure}

\indent Fig.~\ref{fig:Loss} presents the training and validation losses of both the MLP and CNN models, measured in terms of the mean-squared-error (MSE), versus the epoch index, for the two considered datasets, and for the cases when only the LOS component of the channel gain is considered and when both the LOS and NLOS components are considered. In total, 30 epochs have been used for training and validating each model. Specifically, the portion of data advocated for training the ANN models is in fact divided into two subsets, one for training the models to obtain the weights $\mathcal{P}$ and one for validating the generalization error of the obtained weights on the unseen data. Hence, each epoch is a pass through the entire training set in one time. As it can be seen in Fig.~\ref{fig:Loss}, the training and validation losses are decreasing as the epoch index increases which demonstrates that the obtained ANN models are not overfitting and can generalize well over unseen data points in the online testing. \\
\indent Table \ref{T4} presents the average estimation error and the precision of the 3D position and the orientation angles of the UE in the online phase using the proposed CNN and the MLP models and the KNN technique (after the training and the validation) for the two considered datasets, and for the cases when only the LOS component of the channel gain is considered and when both the LOS and NLOS components are considered. This table shows that the CNN model outperforms both the MLP model and the KNN technique. In fact, when considering the total channel gain and a dataset of size $N=10^6$, the proposed CNN model is able to achieve an average positioning error of $10.53$ cm with $90\%$ of the positioning errors below $17.15$ cm (precision), without any prior knowledge on the UE position and orientation and any assumptions on the LiFi system. The same performance and observations can be seen in the average estimation error and the precision of the orientation angles.\\ 
\begin{figure}[t]
		\centering
		\begin{subfigure}[b]{0.9\columnwidth}
			\centering
			\includegraphics[width=\columnwidth,draft=false]{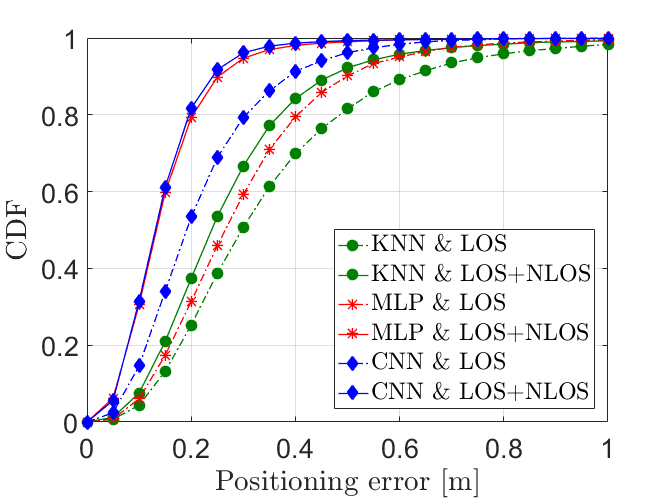}
			\caption{$N = 10^5$}
		\end{subfigure}
		\begin{subfigure}[b]{0.9\columnwidth}
			\centering
			\includegraphics[width=\columnwidth,draft=false]{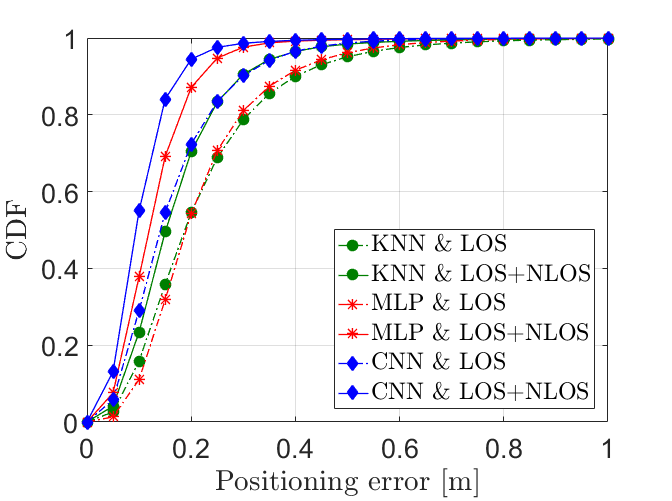}
			\caption{$N = 10^6$}
		\end{subfigure}
		\caption{CDF of the positioning error for the proposed CNN and MLP models with the KNN technique.}
		\label{fig:CNNvsMLPvsKNN_position}
\end{figure}
\begin{figure}[t]
		\centering
		\begin{subfigure}[b]{0.9\columnwidth}
			\centering
			\includegraphics[width=\columnwidth,draft=false]{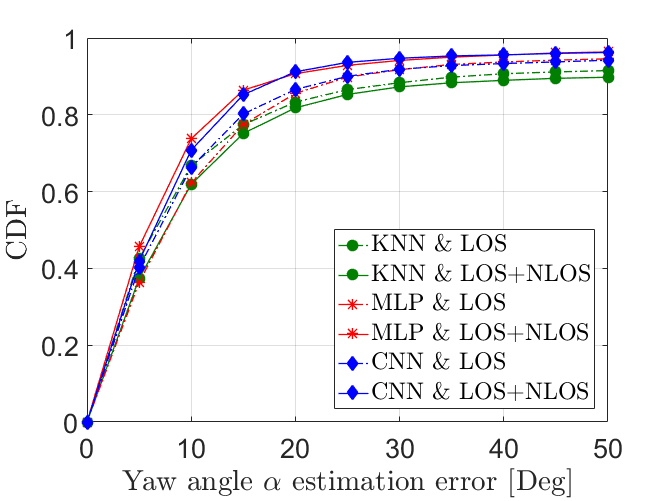}
			\caption{$N = 10^5$}
		\end{subfigure}
		\begin{subfigure}[b]{0.9\columnwidth}
			\centering
			\includegraphics[width=\columnwidth,draft=false]{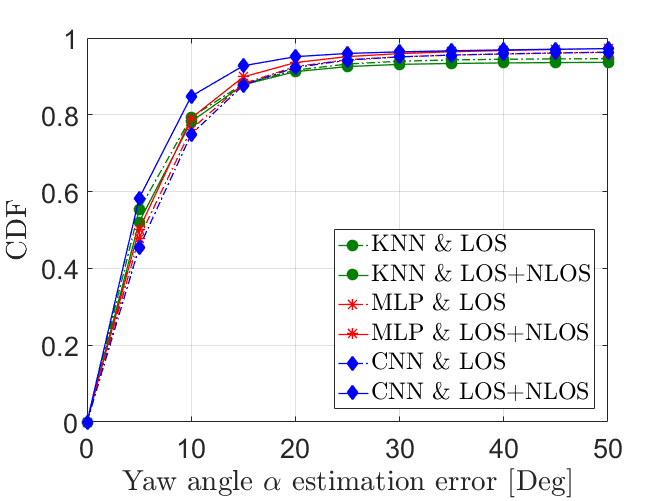}
			\caption{$N = 10^6$}
		\end{subfigure}
		\caption{CDF of the estimation error of the yaw angle $\alpha$. for the proposed CNN and MLP models and the KNN technique.}
		\label{fig:CNNvsMLPvsKNN_alpha}
\end{figure}
\begin{figure}[t]
		\centering
		\begin{subfigure}[b]{0.9\columnwidth}
			\centering
			\includegraphics[width=\columnwidth,draft=false]{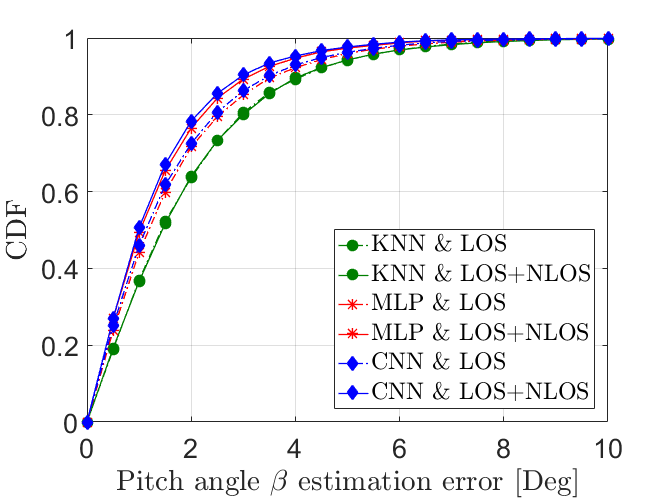}
			\caption{Pitch angle $\beta, N = 10^5$.}
		\end{subfigure}\\
		\begin{subfigure}[b]{0.9\columnwidth}
			\centering
			\includegraphics[width=\columnwidth,draft=false]{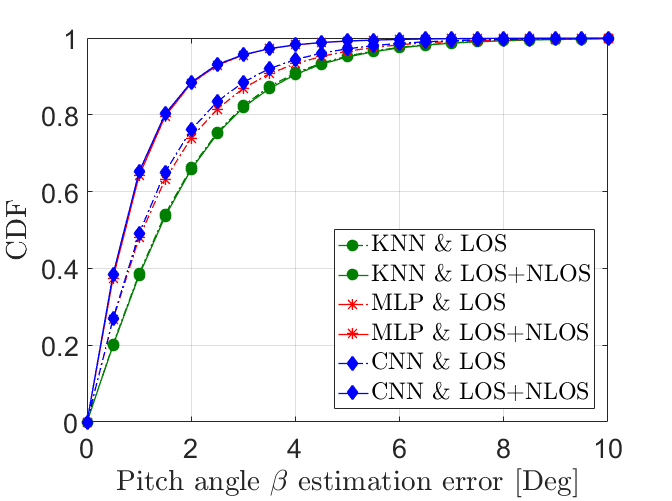}
			\caption{Pitch angle $\beta, N = 10^6$.}
		\end{subfigure}
		\caption{CDF of the estimation error of the pitch angle $\beta$. for the proposed CNN and MLP models and the KNN technique.}
		\label{fig:CNNvsMLPvsKNN_beta}
	\end{figure}
\begin{figure}[t]
		\centering
		\begin{subfigure}[b]{0.9\columnwidth}
			\centering
			\includegraphics[width=\columnwidth,draft=false]{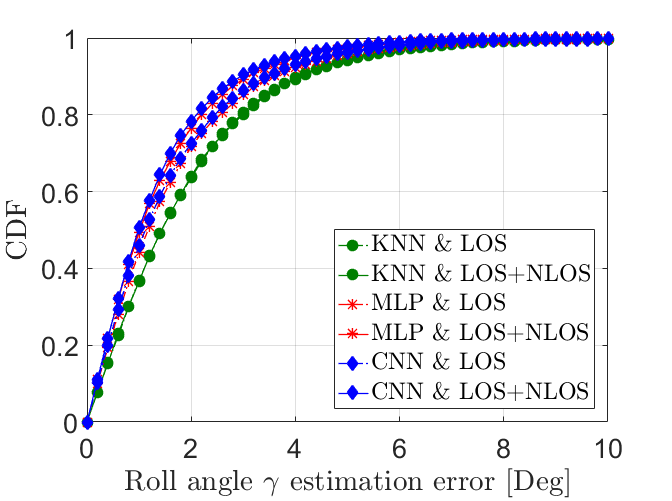}
			\caption{Roll angle $\gamma, N = 10^5$.}
		\end{subfigure}\\
		\begin{subfigure}[b]{0.9\columnwidth}
			\centering
			\includegraphics[width=\columnwidth,draft=false]{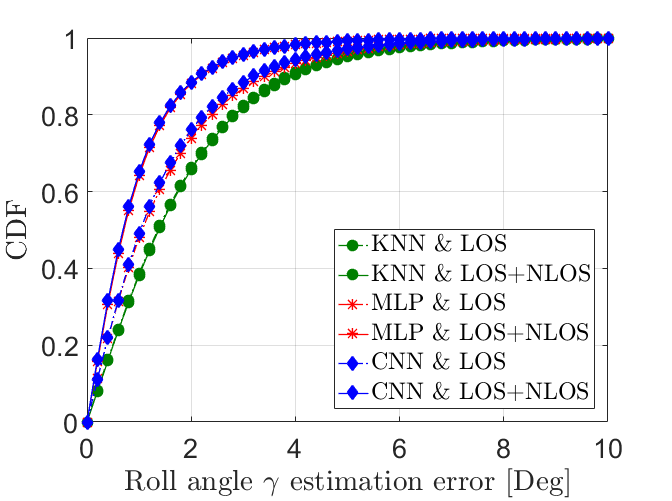}
			\caption{Roll angle $\gamma, N = 10^6$.}
		\end{subfigure}
		\caption{CDF of the estimation error of the roll angle $\gamma$. for the proposed CNN and MLP models and the KNN technique.}
		\label{fig:CNNvsMLPvsKNN_gamma}
	\end{figure}
\indent Figs.~\ref{fig:CNNvsMLPvsKNN_position}, \ref{fig:CNNvsMLPvsKNN_alpha}, \ref{fig:CNNvsMLPvsKNN_beta} and \ref{fig:CNNvsMLPvsKNN_gamma} present the CDF of the instantaneous positioning error and the instantaneous estimation error of the yaw angle $\alpha$, the pitch angle $\beta$ and the roll angle $\gamma$ resulting from the proposed CNN and MLP models and from the KNN technique, for the two considered datasets and for the cases when only the LOS component of the channel gain is considered and when both the LOS and NLOS components are considered. These figures also confirm that the proposed CNN and MLP models outperform the KNN technique, with the CNN being the best model to adopt. Furthermore, jointly with Table \ref{T4}, Figs.~\ref{fig:CNNvsMLPvsKNN_position}, \ref{fig:CNNvsMLPvsKNN_alpha}, \ref{fig:CNNvsMLPvsKNN_beta} and \ref{fig:CNNvsMLPvsKNN_gamma} show that increasing the dataset size increases the learning efficiency of the ANN models. This is mainly due to the fact that having more data points will allow the ANNs to learn better the random behaviour of the environment, which is translated in terms if the effects of the random position and the random orientation of the UE on the instantaneous received SNR. In addition, it can be seen that the estimation performance of the ANN models increases when the total channel gain is considered instead of only the LOS components. This confirms that the NLOS components provide useful information that improves the estimation performance of the 3D position and the orientation angles of the UE, rather than a source of noise as was considered in the literature \cite{zhou2019joint}. In fact, including the NLOS components of a UE with random orientation leads to a non-symmetric SNR change within the room, which is not the case when only the LOS components and upward PD/LED are considered, in which case many positions may have similar SNRs. \\
\indent Another observation that can be seen from Table \ref{T4} and Figs.~\ref{fig:CNNvsMLPvsKNN_position}, \ref{fig:CNNvsMLPvsKNN_alpha}, \ref{fig:CNNvsMLPvsKNN_beta}, \ref{fig:CNNvsMLPvsKNN_gamma} is that the estimation performance of the CNN, the MLP and the KNN models are more accurate for some parameters than others. In fact, we can see that the estimation performance of the pitch angle $\beta$ and the roll angle $\gamma$ is way better than the one of the yaw angle $\alpha$. This is mainly due to the statistics of these angles. Specifically, as shown in Table \ref{distfit}, the standard deviations of $\alpha$, $\beta$ and $\gamma$ are very small. Therefore, when generating the dataset, the random realizations of these angles will fluctuate slightly over their individual means. However, the mean of the yaw angle $\alpha$, which is $\Omega-90^\circ$, is also randomly changing, whereas the means of the pitch angle $\beta$ and the roll angle $\gamma$ are fixed. This explains why the estimation accuracy of the pitch and the roll angles is better than the one of the yaw angle.
\subsection{Reliability Performance Evaluation}
\begin{figure}[t]
\centering     
\includegraphics[width=0.9\linewidth]{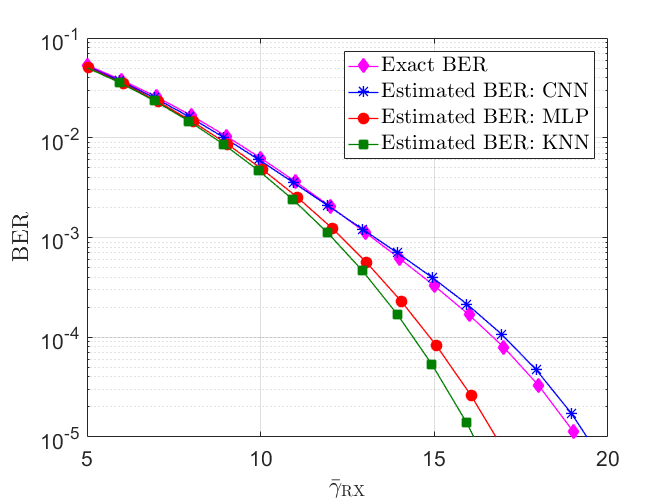}
\caption{Downlink BER evaluation versus the average received SNR $\bar{\gamma}_{\rm RX}$, for an indoor environment with dimensions $L \times W \times H = 5 \times 5 \times 3$ $m^3$ with the number of APs $N_{\rm r} = 16$ and a dataset with size $N = 10^5$.}
\label{fig:BER}
\end{figure}
\begin{table*}[t]
\caption{Computational time per data point [ms]}
\centering
\renewcommand{\arraystretch}{1} 
\setlength{\tabcolsep}{0.24cm} 
\begin{tabular}{| c | c | c | c | c | c | c | c |}
  \cline{3-8} 
  \multicolumn{2}{c}{} & \multicolumn{2}{|c|}{CNN} & \multicolumn{2}{c|}{MLP} & \multicolumn{2}{c|}{KNN} \\ 
  \cline{3-8} 
    \multicolumn{2}{c|}{} & $N=10^5$ & $N=10^6$ & $N=10^5$ & $N=10^6$ & $N=10^5$ & $N=10^6$ \\
    \hline
    \multirow{2}{*}{LOS} & Off-line phase (training phase) & $18$ & $17$ & $2.5$ & $2$ & $3.4 \times 10^{-3}$ & $5.48 \times 10^{-3}$ \\
    \cline{2-8} 
    & On-line phase (testing phase) & $0.19$ & $0.14 $ & $0.09$ & $0.03 $ & $0.01$ & $0.29$ \\ 
    \hline 
     \multirow{2}{*}{NLOS} & Off-line phase (training phase) & 18 & 17 & 2.2 & 2.3 & $3.044\times 10^{-3}$ & $6.94 \times 10^{-3}$ \\
    \cline{2-8} 
    & On-line phase (testing phase) & $0.2$ & $0.2$ & $0.08 $& $0.03$ & $0.01$ & $0.26$ \\ 
    \hline
\end{tabular} 
\label{T5}
\end{table*}
One main objective behind estimating the UE position and orientation, other than the purpose of navigation-based or location-aware services, is estimating the indoor channel gain between the APs and the UE. However, although some navigation-based or location-aware services may tolerate a range of estimation error, it is not the case when estimating the UE channel gain. In fact, the position and orientation estimation errors will affect to the channel estimation accuracy, which in turn will affect the performance of LiFi systems in terms of reliability and achievable rates. Due to this reason, the reliability performance of LiFi systems under the position and orientation estimation errors is investigates in this section. Let $\textbf{h}_d$ denote the $N_{\rm r} \times 1$ total downlink channel gain vector between the APs and the UE, $P_{\rm elec}^{\rm d}$ denotes the electrical transmit power from the APs and $\sigma_{d}^2$ denotes the average noise power at the UE. In this case, the instantaneous received SNR at the UE is given by $\gamma_{\rm RX} = \frac{P_{\rm elec}^{\rm d} || \textbf{h}_{\rm d} ||_1^2}{\sigma_{\rm d}^2}$. Assuming that the on-off keying (OOK) modulation is employed and that the APs broadcast the same signal to the UE, Fig.~\ref{fig:BER} presents the average exact and estimated BERs versus the average received SNR $\bar{\gamma}_{\rm RX} = \mathbb{E} \left[  \gamma_{\rm RX} \right]$ for an indoor environment with dimensions $L \times W \times H = 5 \times 5 \times 3$ $m^3$ and for the number of APs $N_{\rm r} = 16$. The average here is w.r.t to all the realizations of the channel gain vector $\textbf{h}_{\rm d}$ induced by all the exact and the estimated positions and orientations of the UE. \\ 
\indent From Fig.~\ref{fig:BER}, the following observations can be highlighted. First, the proposed CNN model provides the most accurate estimation of the BER compared to the proposed MLP model and the KNN technique. This is mainly due to the fact that, based on the results of the previous subsections, the CNN was shown to be the best model in estimating the position and the orientation of the UE. Second, the proposed MLP model along with the KNN technique provide an underestimation of the BER. In fact, One might think that the estimated average BER by the MLP model and the KNN technique is better than the exact one but it is not what it looks like. In fact, this underestimation is mainly due to the position and orientation estimation errors. Specifically, the estimated position and orientation, although they produce better BER performance, but they don't reflect the realistic BER resulting from the exact position and orientation of the UE.
\subsection{Computational Complexity Evaluation}
\indent Considering the CNN model, the training time per data point is approximately $18$ ms. Thus, the total training time for a dataset with size $N = 10^5$ is approximately equal to $30$ mins and the total training time for a dataset with size $N=10^6$ is approximately equal to $5$ hours. In addition, the required time for generating the dataset with size with size $N = 10^5$ is approximately $1$ hour and 30 mins and the required time for generating the dataset with size $N = 10^6$ is approximately equal to $15$ hours. Although time required for the dataset generation and the models training is large, this high computational complexity is not an issue, since the dataset generation and the models training is performed in the offline phase and the most important computational complexity is the one of the online phase.  \\
\indent Concerning the complexity of the proposed CNN and MLP models, it can be seen that computational time in the on-line phase is extremely low. In fact, for the two considered datasets, the maximum average computational time in the on-line phase is approximately $0.19$ ms, i.e., real time estimation. Nevertheless, the best existing joint position and orientation estimation technique, which was proposed in \cite{zhou2019joint}, has an average computational time of $0.5$ s, which demonstrates the potential of the proposed ANN models in providing highly accurate and real-time joint position and orientation estimation.
\subsection{Effects of the Indoor Environment Geometry}
\begin{figure}[t]
\centering     
\includegraphics[width=0.9\linewidth]{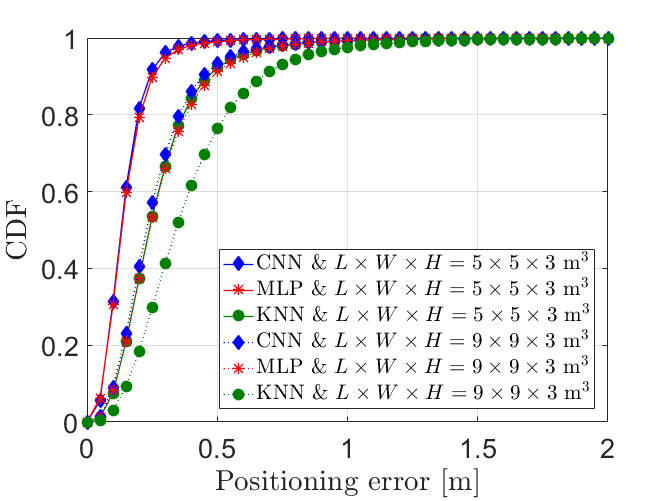}
\caption{CDF of the positioning error for the proposed CNN and MLP models with the KNN technique for different room dimensions $L \times W \times H$, where the number of APs $N_{\rm r} = 16$ and the total channel gain is considered}
\label{fig:dimension}
\end{figure}
\begin{figure}[t]
\centering     
\includegraphics[width=0.9\linewidth]{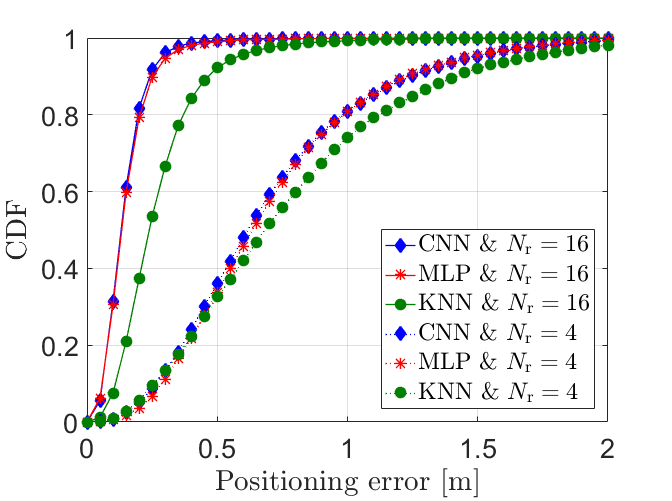}
\caption{CDF of the positioning error for the proposed CNN and MLP models with the KNN technique for different number of APs $N_{\rm r}$, where the room dimension is $L \times W \times H = 5 \times 5 \times 3$ m$^3$ and the total channel gain is considered.}
\label{fig:APs}
\end{figure}

Fig.~\ref{fig:dimension} presents the CDF of the positioning error for the proposed CNN and MLP models and the KNN technique for different room dimensions $L \times W \times H$, where the number of APs $N_{\rm r} = 16$ and the total channel gain is considered. This figure shows that increasing the room dimension while keeping the number of APs fixed will decrease the estimation performance. This observation is expected, since increasing the room dimensions will enlarge the search space of the unknown position parameters, and therefore, the probability of wrong estimation will increase. \\ 
\indent Fig.~\ref{fig:APs} presents the CDF of the positioning error for the proposed CNN and MLP models and the KNN technique for different number of APs $N_{\rm r}$, where the room dimension is $L \times W \times H = 5 \times 5 \times 3$ m$^3$ and the total channel gain is considered. This figure shows that decreasing the number of APs while keeping the room dimension fixed will decrease the estimation performance. This observation is also expected since decreasing the number of APs will decrease the number of received SNR values, and therefore, the size of the feature SNR vector will decrease, which reduces the amount of information that can be exploited for the estimation.
\section{Conclusions and Future Work}
\begin{figure}[t]
		\centering
	\includegraphics[width=0.7\columnwidth,draft=false]{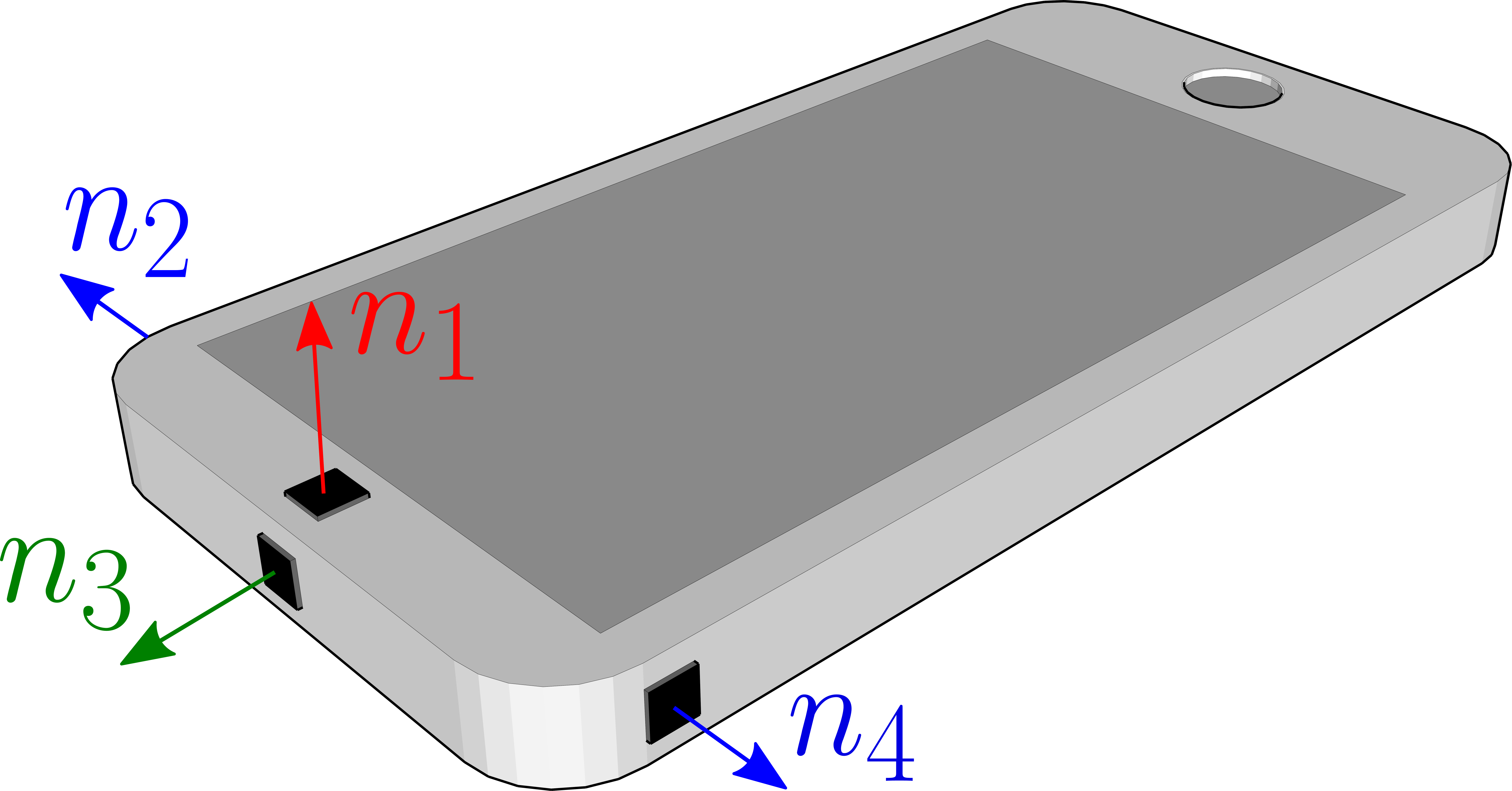}
	\caption{Downlink BER versus the average received SNR at the UE for different numbers of APs $N_{\rm r}$.}
	\label{smartphone2}
\end{figure}

In this paper, novel ANN models were proposed for the joint 3D position and orientation estimation of a randomly located LiFi UE with a random orientation and an unknown emitting power. Using RSS-based fingerprinting, the proposed approach consisted of generating a measurement-based dataset that contains the instantaneous received SNR along with the corresponding 3D position and orientation angles. Then an MLP and a CNN models were designed to map efficiently the SNR feature vectors with the corresponding positions and orientation angles. Although there were no prior knowledge on the UE position and orientation and any assumptions on the LiFi system, the proposed models were able to achieve centimeter level positioning error and high accurate orientation angles estimation. The performance of the proposed models were compared with the KNN technique, in terms of average estimation error, precision, BER and computational time, where the superiority of the proposed ANN models was shown. These results have enlightened the potential of the proposed ANN models in providing highly accurate and real-time joint position and orientation estimation. In addition, and against the existing position and orientation solutions, the results have demonstrated how including the NLOS components of the channel gain can improve significantly the position and orientation estimation performance. \\ 
\indent As it was mentioned in the paper, the performance of any position and orientation estimation technique depends mainly on the channel matrix, which determines the differentiability between the received signal vectors. A promising idea to improve more the performance of the proposed models is by exploiting the spatial diversity at the UE. In fact, by using only one transmit IR-LED at the UE, the channel gain is very susceptible to indoor blockage effects. In addition, it is highly likely that the IR-LED is out of the FOV of all PDs of the APs since usually the smart phone is held with an orientation other than upward. Therefore,
exploiting the ``multi-directional transmitter'' (MDT) design of the UE that was proposed in \cite{mohammad2018optical} is indeed an interesting future research direction. The MDT configuration consists of placing IR-LEDs on the screen and three other sides of the mobile device, as shown in Fig. \ref{smartphone2}. Note that another LED can be placed at the back, which can be activated instead of the one on the screen for situations where the user is placed on a horizontal surface. Such configuration can overcome the effects of random orientation and link blockage, and hence, it is indeed a promising solution for estimating accurately the position and the orientation of LiFi devices.
\section*{Acknowledgment}
M. A. Arfaoui and C. Assi acknowledge the financial support from the Natural Sciences and Engineering Research Council of Canada (NSERC), Fonds Quebecois de la Recherche sur la Nature et les Technologies (FQRNT) and from Concordia University. 
A. Ghrayeb is supported in part by Qatar National Research Fund under NPRP Grant NPRP8-052-2-029 and in part by FQRNT. M. D. Soltani and M. Safari gratefully acknowledge Engineering and Physical Sciences Research Council (EPSRC) under grant EP/S016570/1 `Terabit Bidirectional Multi-User Optical Wireless System (TOWS) for 6G LiFi'.
H. Haas acknowledges the financial support from the Wolfson Foundation and Royal Society. He also gratefully acknowledges financial support by the EPSRC under the Established Career Fellowship grants EP/R007101/1 and EP/S016570/1 `Terabit Bidirectional Multi-User Optical Wireless System (TOWS) for 6G LiFi'. 
\appendices 
\bibliographystyle{IEEEtran}
\bibliography{main.bib}
\end{document}